\documentclass[aps,prd,a4paper,twocolumn,amsmath,showpacs,superscriptaddress,nofootinbib,preprintnumbers]{revtex4-1}

\usepackage{verbatim}
\usepackage{epsfig}
\usepackage{graphicx}
\usepackage{multirow}
\usepackage{dcolumn}
\usepackage{amsmath}
\usepackage{mathtools}
\usepackage{amsfonts}
\usepackage{amssymb}
\usepackage{epstopdf}
\usepackage{bm}
\usepackage{siunitx}
\usepackage{braket}
\usepackage{enumitem}
\usepackage{soul}
\usepackage[table]{xcolor}
\usepackage{colortbl}

\definecolor{navyblue}{rgb}{0.0, 0.0, 0.5}
\definecolor{royalblue}{rgb}{0.25, 0.41, 0.88}
\definecolor{cadmiumgreen}{rgb}{0.0, 0.42, 0.24}
\definecolor{blue-violet}{rgb}{0.54, 0.17, 0.89}
\definecolor{darkviolet}{rgb}{0.58, 0.0, 0.83}
\definecolor{orange(colorwheel)}{rgb}{1.0, 0.5, 0.0}
\definecolor{Gray}{gray}{0.9}
\definecolor{LightCyan}{rgb}{0.88,1,1}

\usepackage{hyperref}
\hypersetup{
    colorlinks=true, 
    linkcolor=royalblue, 
    citecolor=magenta}

\newcommand\ee{\end{equation}}
\newcommand\be{\begin{equation}}
\newcommand\eea{\end{eqnarray}}
\newcommand\bea{\begin{eqnarray}}

\newcommand{\ns}{n_{\rm s}}

\newcommand\limit[1]{#1\%\,\mathrm{CL}}

\newcommand{\tblack}[1]{\textcolor{black}{#1}} 

\newcommand{\tu}{\textup}
\newcommand{\dif}{\mathrm{d}}
\newcommand\ie{{\it i.e.}~}
\newcommand\eg{{\it e.g.}~}

\newcommand\eq[1]{Eq.~\eqref{eq:#1}}
\newcommand\sect[1]{Sec.~\ref{sec:#1}}
\newcommand\fig[1]{Fig.~\ref{fig:#1}}
\newcommand\tab[1]{Tab.~\ref{tab:#1}}

\renewcommand{\vec}{\bm}
\newcommand{\mufid}{\mu^{(\text{fid})}}
\newcommand{\nrun}{\alpha_\mathrm{s}}
\newcommand{\nrunfid}{\alpha_\mathrm{s}^{(\text{fid})}}

\newcommand\vertsp{\rule[-2mm]{1mm}{0mm} &}
\newcommand\horsp{\rule[-2mm]{0mm}{5.5mm}}

\DeclarePairedDelimiter{\abs}{\lvert}{\rvert}

\begin{document}

\title{$\mu$-Distortions or Running: A Guaranteed Discovery from CMB Spectrometry
}

\author{Giovanni Cabass}
\affiliation{Physics Department and INFN, Universit\`a di Roma 
	``La Sapienza'', P.le\ Aldo Moro 2, 00185, Rome, Italy}
	
\author{Alessandro Melchiorri}
\affiliation{Physics Department and INFN, Universit\`a di Roma 
	``La Sapienza'', P.le\ Aldo Moro 2, 00185, Rome, Italy}
	
\author{Enrico Pajer}
\affiliation{Institute for Theoretical Physics and Center for Extreme Matter and Emergent Phenomena,
	Utrecht University, Leuvenlaan 4, 3584 CE Utrecht, The Netherlands}

\begin{abstract}
\noindent We discuss the implications 
of a PIXIE-like experiment, which would measure $\mu$-type spectral distortions of the CMB 
at a level of $\sigma_{\mu}=(1/n)\times 10^{-8}$, with $n\geq1$ representing an improved sensitivity (\textit{e.g.} $n=10$ corresponds to PRISM). Using \textit{Planck} data and considering the six-parameter $\Lambda$CDM model, we compute the posterior for $\mu_8\equiv\mu\times 10^{8}$ and find $\mu_8=1.57^{+0.11}_{-0.13}$ ($68\%\,\mathrm{CL}$). This becomes $\mu_{8}= 1.28^{+0.30}_{-0.52}$ ($68\%\,\mathrm{CL}$) when the running $\alpha_\mathrm{s}$ of the spectral index is included. 
We point out that a sensitivity of about $3\times$ PIXIE implies a \textit{guaranteed discovery}: $\mu$-distortion is detected or $\alpha_\mathrm{s}\geq 0$ is excluded (both at $95\%\,\mathrm{CL}$ or higher). This threshold sensitivity sets a clear benchmark for CMB spectrometry. 
For a combined analysis of PIXIE and current \textit{Planck} data, we discuss the improvement on 
measurements of the tilt $n_\mathrm{s}$ and the running $\alpha_\mathrm{s}$ and the dependence on the choice of the pivot. A fiducial running of $\alpha_\mathrm{s}=-0.01$ (close to the \textit{Planck} best-fit) leads to a detection of negative running at 
$2\sigma$ for $5\times$ PIXIE. A fiducial running of $\alpha_\mathrm{s}=-0.02$, still compatible with \textit{Planck}, requires $3\times$ PIXIE to rule out $\alpha_\mathrm{s} = 0$ (at $95\%\,\mathrm{CL}$). We propose a convenient and compact visualization of the improving constraints on the tilt, running and tensor-to-scalar ratio.
\end{abstract}


\maketitle

\twocolumngrid


\section{Introduction}
\label{sec:introduction}

\noindent The recent measurements of Cosmic Microwave Background (CMB) anisotropies made by the 
\emph{Planck} satellite experiment \cite{Ade:2015lrj} have provided, once again, a 
spectacular confirmation of the $\Lambda$CDM cosmological model 
and determined its parameters with an impressive accuracy. 
Also, numerous new ground based or balloon borne CMB telescopes are currently gathering data or under development. 
Moreover, several proposals for a new satellite experiment like PIXIE \cite{Kogut:2011xw}, 
PRISM \cite{Andre:2013afa, Andre:2013nfa}, CORE \cite{Bouchet:2011ck}, and LiteBIRD 
\cite{Matsumura:2013aja} are under discussion.

In summary, two main lines of investigation are currently pursued: CMB polarization and spectral distortions. 
Improving current measurements of CMB polarization is partially motivated by the inflationary paradigm. 
As it is well known, the simplest models of inflation predict a nearly scale-invariant (red-tilted) spectrum of primordial scalar perturbations, in perfect agreement with the latest experimental evidence. Inflation also predicts a stochastic background of gravitational waves: a discovery of this background (\eg through measurements of CMB $B$-mode polarization \cite{Zaldarriaga:1996xe, Kamionkowski:1996ks}) with a tensor-to-scalar ratio $r\sim10^{-2}$ would correspond to inflation occurring at the GUT scale. Planned and/or proposed CMB experiments could detect this background, and measure the tensor-to-scalar ratio $r\sim 0.01\times(E_\tu{inflation}/\num{d16}\,\mathrm{GeV})^4$ with a relative error of order $
\num{d-2}$, if inflation occurs at these energies \cite{Creminelli:2015oda, Cabass:2015jwe}. This would be a spectacular confirmation of the inflationary theory. However, the energy scale of inflation could be orders of magnitude lower than the GUT scale. In this case, the stochastic background would be out of the reach of upcoming or planned experiments.

On the other hand, CMB $\mu$-type spectral distortions are an unavoidable prediction of the $\Lambda$CDM model, since they are generated by the damping of primordial fluctuations \cite{Sunyaev:1970er, Hu:1992dc} with an amplitude of order $\mu = \mathcal{O}(\num{d-8})$ (for this reason, it will be useful to define the parameter $\mu_8\equiv\mu\times\num{d8}$, that will be used in the rest of the paper). 

While a measurement of CMB spectral distortions could shed light on several aspects of physics beyond $\Lambda$CDM such as, \textit{e.g.}, gravitino decay 
\cite{Dimastrogiovanni:2015wvk}, cosmic strings \cite{Anthonisen:2015tra}, magnetic fields \cite{Wagstaff:2015jaa}, hidden photons \cite{Kunze:2015noa}, and dark matter interactions \cite{Ali-Haimoud:2015pwa}, just to name a few, we stress that spectral distortions could provide significant information on inflation 
through the contribution coming from primordial perturbations \cite{Dent:2012ne, Chluba:2012we, Khatri:2013dha, Clesse:2014pna, Enqvist:2015njy}.

Indeed, in a typical inflationary model, the spectral index $\ns$ of scalar perturbations is expected to have a small (and often negative) running, 
of order $\abs{\nrun}\sim (1-\ns)^{2}$ \cite{Roest:2013fha, Garcia-Bellido:2014gna, Gobbetti:2015cya}. 
State-of-the art CMB observations by the \emph{Planck} experiment \cite{Ade:2015lrj, Ade:2015xua} are fully compatible with an exact power law spectrum of primordial fluctuations $P_\zeta(k)$, with $\nrun = -0.006\pm0.007$ at $\limit{68}$ (\emph{Planck} $TT$, $TE$, $EE$ + lowP dataset). A more than ten-fold improvement in sensitivity is therefore needed to reach the typical slow-roll values with CMB experiments. However, CMB anisotropies can probe $P_\zeta(k)$ only up to $k\approx\num{0.1}\,\mathrm{Mpc}^{-1}$, since at shorter scales primordial anisotropies are washed away by Silk damping \cite{Silk:1967kq, Peebles:1970ag, Kaiser:1983abc} and foregrounds become dominant. There is, then, a limit in the range of multipoles that we can use to test the scale dependence of the power spectrum.\footnote{For this reason, we expect that $E$-mode polarization will be better, in the long run, at constraining the scale dependence of $P_\zeta(k)$, since $C^{EE}_\ell$ starts to become damped around $\ell\approx2500$ (see \cite{Galli:2014kla} for a discussion).} Moreover, CMB measurements will soon be limited by cosmic variance: recent analyses have shown that for upcoming experiments (COrE+ or CMB Stage IV), which are close to be CVL (Cosmic Variance Limited), one can expect $\sigma_{\nrun}\approx\num{d-3}$ \cite{Wu:2014hta, Dore:2014cca, Errard:2015cxa}. 

The CMB $\mu$-type spectral distortion is sensitive to the amount of scalar power up to $k$ of order $\num{d4}\,\mathrm{Mpc}^{-1}$ because of the damping of acoustic modes. The strong lever arm makes this observable an ideal probe to improve the bounds on the running from large scale CMB anisotropies. In addition, the cosmic variance of the $\mu$ monopole and of the higher multipoles is minuscule (see \cite{Pajer:2012vz} 
for a discussion). With a sufficiently broad frequency coverage, instrumental noise will be the main source of uncertainty for any foreseeable future, leaving ample room for improvements.

In this context, we address several questions: 
\begin{itemize}[leftmargin=*]
\item is there a benchmark sensitivity for CMB spectrometry, \ie which should be the target of the next generation experiments? How can we design an experiment to ensure a discovery even in the absence of a detection?
\item what sensitivity to the spectrum is needed to detect $\mu$-distortions when accounting for the prior knowledge from \emph{Planck}?
\item how much will a joint analysis of large scale CMB anisotropies and CMB spectral distortion strengthen the bounds on the running? How does this quantitatively depend on the improvement over PIXIE sensitivity?\footnote{For example the PRISM imager \cite{Andre:2013afa, Andre:2013nfa} corresponds to approximately $10\times$ PIXIE.} 
\end{itemize}

To articulate the answers to these questions, we consider the following three fiducial cosmologies: 
\begin{itemize}[leftmargin=*]
\item {a $\Lambda\mathrm{CDM}$ cosmology with zero running: the best-fit for the $\mu$-amplitude, in this case, is of order $\mu_8 = \num{1.6}$.} We stress that for the sensitivities considered in this work, this fiducial is indistinguishable from models with running of order $(1-\ns)^2$, such as typical slow-roll models;
\item {a fiducial 
spectral distortion amplitude $\mufid_8$ equal to the best-fit of the the \emph{Planck} analysis for the $\Lambda\mathrm{CDM} + \nrun$ model, \ie $\mufid_8 = \num{1.06}$.} This value of $\mu$ is roughly correspondent to what one obtains for a running $\nrun = -0.01$ which is close to the mean value predicted by current \emph{Planck} data; 
\item $\nrunfid = -0.02$ (corresponding to $\mu_8 = \num{0.73}$), at the edge of the $2\sigma$ bounds of \emph{Planck}. We note that it is possible to obtain such large negative runnings in some models of single-field inflation like, \textit{e.g.}, extra-dimensional versions of Natural Inflation \cite{ArkaniHamed:2003wu, Feng:2003mk} or recent developments in axion monodromy inflation \cite{Kobayashi:2010pz, Jain:2015jpa, Parameswaran:2016qqq}.
\end{itemize}

The paper is organized as follows: after a brief review of photon thermodynamics in the early universe 
and of distortions from Silk damping (
\sect{distortions-review}), we compute the $\mu$-distortion parameter allowed by current \emph{Planck} bounds for a $\Lambda\text{CDM}$ and $\Lambda\text{CDM} + \nrun$ model (\sect{expectations}). We then analyze what a PIXIE-like 
mission will be able to say about the running, given these posteriors for $\mu$. The discussion is divided in three sections: we start with the predicted bounds on $\mu$-distortions from current \emph{Planck} data (\sect{expectations}). We proceed with a Fisher analysis (\sect{fisher}), discussing also the optimal choice of pivot scale for a combined study of CMB anisotropies and spectral distortions. The MCMC analysis and forecasts are carried out in \sect{mcmc}. Finally, \sect{slow-roll} studies the implications of these results for single-clock slow-roll inflation, and we draw our conclusions in \sect{conclusion}.


\section{Photon thermodynamics}
\label{sec:distortions-review}

\noindent At very early times, for redshifts larger than $z_\tu{dC}\approx\num{2d6}$, processes like double Compton scattering and bremsstrahlung are very efficient and maintain thermodynamic equilibrium: any perturbation to the system is thermalized and the spectrum of the CMB is given to very high accuracy by a black-body. At later times the photon number is effectively frozen, since photons can be created at low frequencies by elastic Compton scattering but their re-scattering at high frequencies via double Compton scattering and bremsstrahlung is not efficient due to the expansion of the universe \cite{Sunyaev:1970er, Ilia1975, DandD1982, Buri1991, Chluba:2011hw, Khatri:2012tv}. 

The end result is a Bose-Einstein distribution $1/(e^{x + \mu(x)} - 1)$ ($x\equiv h\nu/k_\tu{B}T$) with chemical potential $\mu$. Since photons can still be created at low frequencies, $\mu$ will not exactly be frequency independent: it can be approximated as $\mu_\infty \exp(-x_c/x)$, with $x_c\approx\num{5d-3}$. However, no planned/proposed experiments will be able to probe such low frequencies: for this reason we will take the chemical potential to be a constant (and drop the subscript $\infty$). 

For a given energy release $\dif (Q/\rho_\gamma)/\dif z$, one can write the value of $\mu$ as (see \sect{appendix-release}) 
\begin{equation}
\label{eq:mu-general}
\mu(z) = 1.4\int_z^{z_\tu{dC}}\dif z'\frac{\dif (Q/\rho_\gamma)}{\dif z'} e^{-\tau_\tu{dC}(z')}\,\,,
\end{equation}
where the \emph{distortion visibility function} $\tau_\tu{dC}(z)$ can be approximated as $(z/z_\tu{dC})^{5/2}$ \cite{Sunyaev:1970er, Ilia1975, DandD1982, Buri1991, Chluba:2011hw, Khatri:2012tv, Sunyaev:2013aoa}.

Below redshifts around $z = z_{\mu\text{-}i}\approx\num{2d5}$, Compton scattering is not sufficient to maintain a Bose-Einstein spectrum in the presence of energy injection. The distortions generated will then be neither of the $\mu$-type nor of the $y$-type: they will depend on the redshift at which energy injection occurs \cite{Chluba:2011hw, Khatri:2012tw, Khatri:2013dha}, and must be calculated numerically by solving the Boltzmann equation (known as Kompaneet’s equation \cite{kompaneets}, when restricted to Compton scattering). Recently, in \cite{Khatri:2012tw, Khatri:2013dha}, a set of Green's functions for the computation of these intermediate distortions has been provided:\footnote{\tblack{We refer also to \cite{Chluba:2013vsa} for an alternate derivation.}} they sample the intermediate photon spectrum $n^{(i)}$ for a energy release $Q_\tu{ref}/\rho_\gamma = \num{4d-5}$ in $\mathcal{O}(\num{d3})$ redshift bins from $z \approx\num{2d5}$ to $z \approx\num{1.5d4}$. 
The $i$-type occupation number, 
for a generic energy injection history $\dif (Q/\rho_\gamma)/\dif z$, will then be computed as \cite{Khatri:2013dha}
\begin{equation}
\label{eq:i-general}
\begin{split}
I^{(i)}(\nu) &= \frac{2h\nu^3}{c^2}\sum_{z_k} \frac{n^{(i)}_{z_k}(\nu)}{\num{4d-5}}\frac{\dif (Q/\rho_\gamma)}{\dif z}\bigg|_{z_k}\delta z_k \\
&\equiv\frac{2h\nu^3}{c^2}\sum_{z_k} \frac{n^{(i)}_{z_k}(\nu)}{\num{4d-5}}\times\mu^{(i)}_{z_k}\,\,. 
\end{split}
\end{equation}

At redshifts $z\lesssim\num{1.5d4}$ also elastic Compton scattering is not efficient enough: there is no kinetic equilibrium and the distortion is of $y$-type. \tblack{The transition between $\mu$- and $y$-distortions can be modeled with a redshift dependent visibility function \cite{Chluba:2013vsa}. The information on the transition is encoded in the residual $r$-type distortions. 
Since 
$r$-distortions are not degenerate with $\mu$- and $y$-type distortions (see \sect{appendix-shapes}), they can be useful for 
probing the redshift dependence of different energy release histories \cite{Khatri:2013dha, Chluba:2013wsa}.}\footnote{\tblack{As \cite{Chluba:2011hw} shows, they can be used to put constraints on observables like the lifetime of decaying dark matter particles.}}

The $y$-type distortions is expected to be dominated by astrophysics at low redshifts (created when the CMB photons are scattered in the clusters of galaxies by hot electrons, the tSZ effect). While this signal is very interesting by itself as a probe of the matter distribution in the universe \cite{Hill:2013baa, Hill:2015tqa, Hill:2015kua}, our goal is studying the contribution due to dissipation of acoustic waves, and so we will marginalize over it in our analysis (see \sect{appendix-shapes}).\footnote{\tblack{We note that in \cite{Chluba:2013pya}, the authors carried out this marginalization by taking into account also $r$-distortions: this results in a slightly higher $\mu$ detection limits, but does not affect the main results of this paper.}}

{Additional spectral distortions are the ones created during recombination }\cite{Sunyaev:2013aoa, Chluba:2015gta}{ and reionization }\cite{Basu:2003th, Sunyaev:2013aoa, DeZotti:2015awh}{. Previous work on recombination spectra has been carried out in }\cite{Peebles:1968ja, Zeldovich:1969en, Dubro1975, Chluba:2005uz, RubinoMartin:2006ug, Chluba:2006bc}{, and recently }\cite{Chluba:2015gta}{ has shown that spectral distortions from recombination can be computed with high }
{precision. Therefore we are not going to include them in our analysis, assuming they can be subtracted when looking for the primordial signal.}

In this work we will not consider these intermediate distortions, and take the transition between the $\mu$ and $y$ era to be instantaneous at a redshift $z_{\mu\text{-}y}\approx\num{5d4}$ \cite{Hu:1992dc}: in the case of an energy release that does not vary abruptly with redshift, we do not expect the inclusion of \tblack{$r$-distortions} to alter significantly the constraints on the parameters describing $\dif (Q/\rho_\gamma)/\dif z$. We leave the analysis of their effect on \tblack{combined CMB anisotropies - CMB distortions} forecasts for cosmological parameters for future work \tblack{(referring to \cite{Khatri:2013dha, Chluba:2013wsa, Chluba:2013pya} for forecasts involving CMB spectrometry alone)}. 



\begin{figure}[!hbt]
\includegraphics[width=0.48\textwidth]{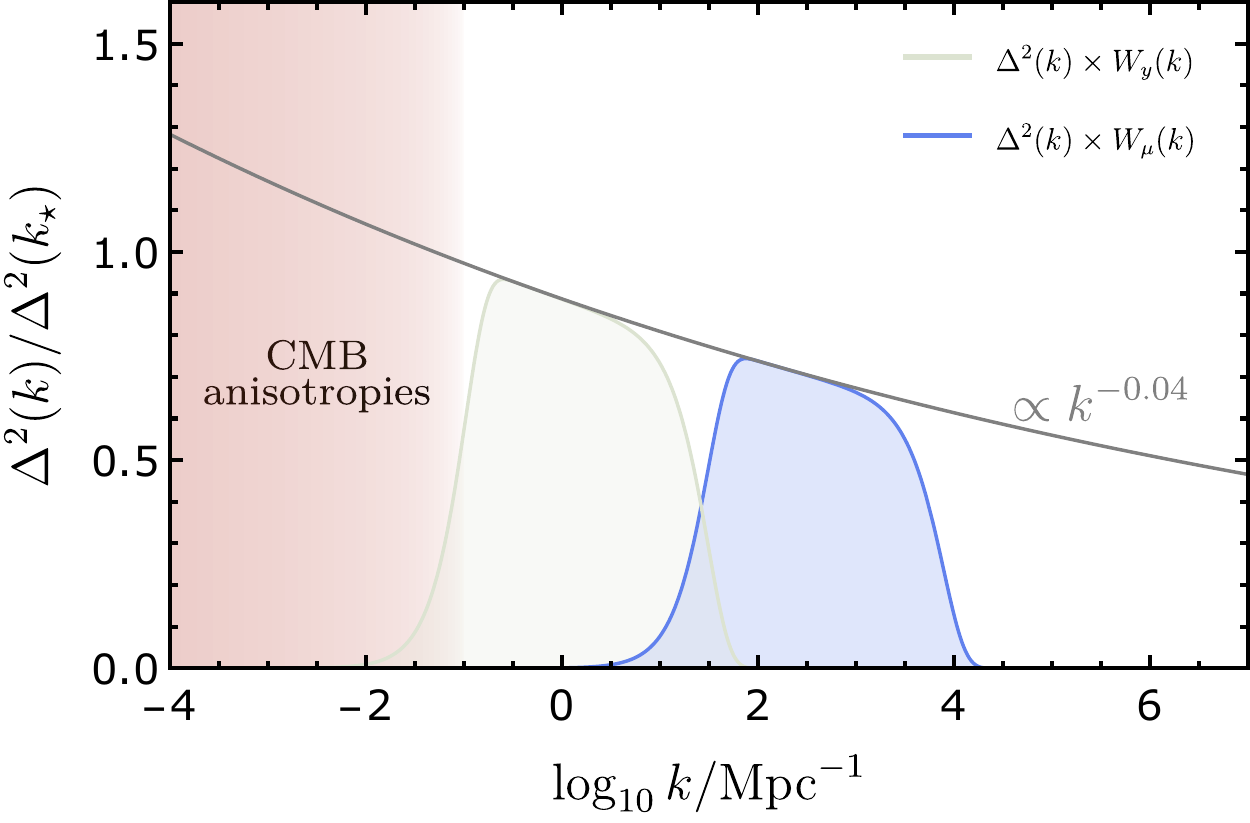}
\caption{\footnotesize{This cartoon plot shows the scales which are probed by $\mu$- 
and $y$-type spectral distortions, using the ``window function'' approximation of \eq{mu-window}.}}
\label{fig:windows}
\end{figure}

While there are many non-standard potential sources of spectral distortions, \eg decaying or annihilating Dark Matter particles \cite{Chluba:2011hw, Khatri:2012tw, Dimastrogiovanni:2015wvk}, a source of heating that is present also in the standard picture is the dissipation of perturbations in the primordial plasma due to Silk damping. Even before recombination, when the tight-coupling approximation holds, photons are random-walking within the plasma with a mean free path $\lambda_\mathrm{mfp} = (n_e\sigma_\tu{T})^{-1}$. In the fluid description, this amounts to anisotropic stresses that induce dissipation. One can compute the (integrated) fractional energy lost by these acoustic waves $\delta_\gamma$: in the tight-coupling approximation \eq{mu-general} reduces to \cite{Chluba:2012gq, Khatri:2012rt}
\begin{equation}
\label{eq:mu-silk}
\begin{split}
\mu 
&\approx\frac{1.4}{4}\braket{\delta_\gamma^2(z,\vec{x})}_p\Big|_{z_{\mu\text{-}y}}^{z_\mathrm{dC}} \\
&\approx 2.3\int\frac{\dif\vec{k}_1 \dif\vec{k}_2}{(2\pi)^3}e^{i\vec{k}_+\cdot\vec{x}}\zeta_{\vec{k}_1}\zeta_{\vec{k}_2} 
e^{-(k^2_1 + k^2_2)/k^2_D}\Big|^{z_\mathrm{dC}}_{z_{\mu\text{-}y}}\,\,, 
\end{split}
\end{equation}
where $\braket{\dots}_p$ indicates the average over a period of oscillation and $\zeta$ is the primordial curvature perturbation. The diffusion damping length appearing in the above formula, instead, is given by \cite{Silk:1967kq, Peebles:1970ag, Kaiser:1983abc}
\begin{equation}
\label{eq:diff-dist}
k_\tu{D}(z) = \sqrt{\int^{+\infty}_z\dif z\frac{1+z}{H n_e\sigma_\tu{T}}\bigg[\frac{R^2 + \frac{16}{15}(1+R)}{6(1+R)^2}\bigg]}\,\,.
\end{equation}
If we consider the ensemble average of $\mu$, we see that it is equal to the log-integral of the primordial power spectrum multiplied by a window function 
\begin{equation}
\label{eq:mu-window}
W_\mu(k) = 2.3\,e^{-2k^2/k^2_D}\Big|^{z_\mathrm{dC}}_{z_{\mu\text{-}y}}\,\,,
\end{equation}
Since the tight-coupling approximation is very accurate at redshifts much before recombination 
we expect this to be a good approximation for the $\mu$-distortion amplitude. This window function and the analogous one for $y$-distortions are shown in \fig{windows}.


This simplified picture allows us to obtain a qualitative understanding of the possible constraints coming from an experiment like PIXIE \cite{Kogut:2011xw}. 

We also account for adiabatic cooling \cite{Chluba:2011hw, Khatri:2011aj}, namely the fact that electrons and baryons alone would cool down faster than photons. Because of the continuous interactions, they effectively extract energy from the photons to maintain the same temperature, leading to an additional source of distortions of the CMB spectrum. During the $\mu$-era, this energy extraction results in a negative $\mu$-distortion of order $\mu_\tu{BEC}\approx\num{-2.7d-9}$ (for the \emph{Planck} 2015 best-fit values of cosmological parameters). 



\section{Expectations from large scales}
\label{sec:expectations}

\begin{figure}[!hbt]
\includegraphics[width=0.48\textwidth]{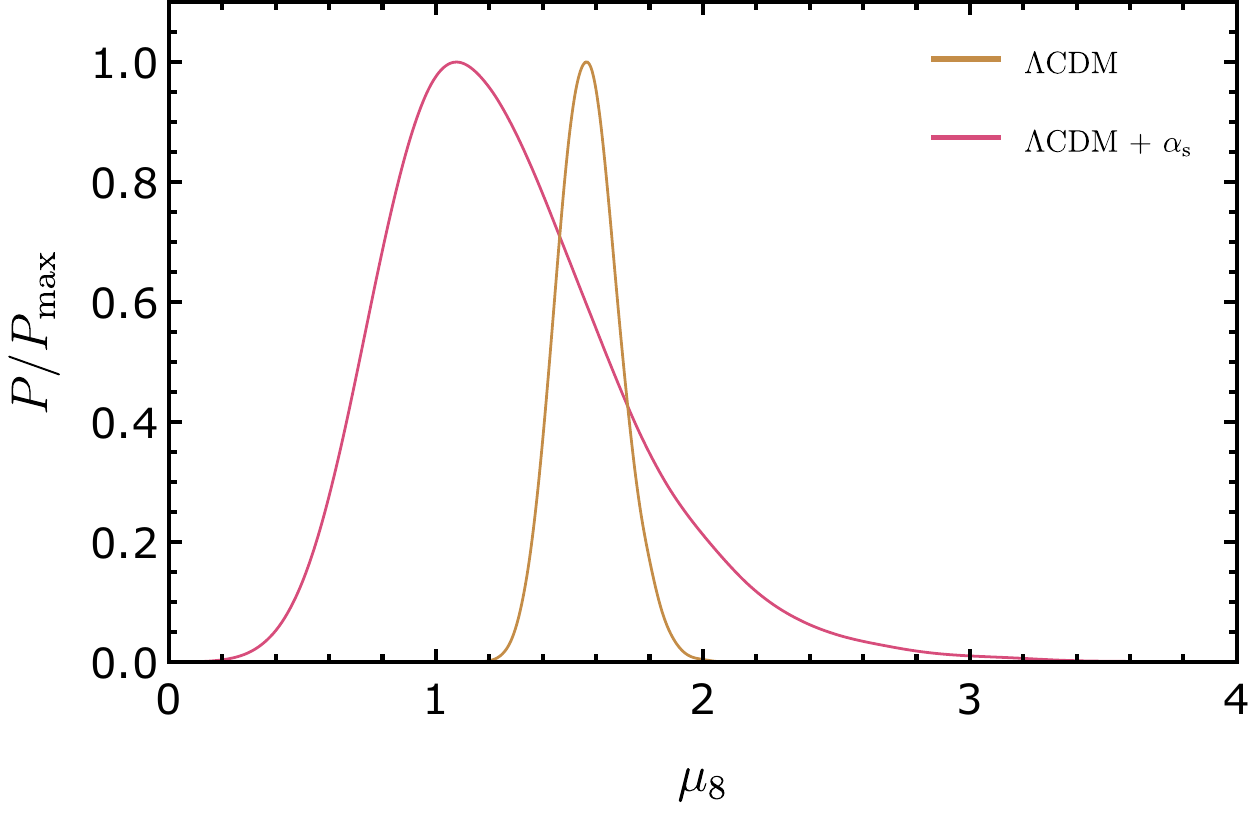}
\caption{\footnotesize{The figure shows the one-dimensional posteriors for $\mu_8$ predicted by \emph{Planck} $TT$, $TE$, $EE$ + lowP data, for the $\Lambda\mathrm{CDM}$ model (orange curve) and the $\Lambda\mathrm{CDM} + \nrun$ model (purple curve)
. The posteriors have been obtained through the ``\texttt{idistort}'' code developed by Khatri and Sunyaev \cite{Khatri:2012tw, Khatri:2013dha}.}}
\label{fig:mu-1d}
\end{figure}


\begin{figure*}
\begin{center}
\begin{tabular}{c c}
\includegraphics[width=\columnwidth]{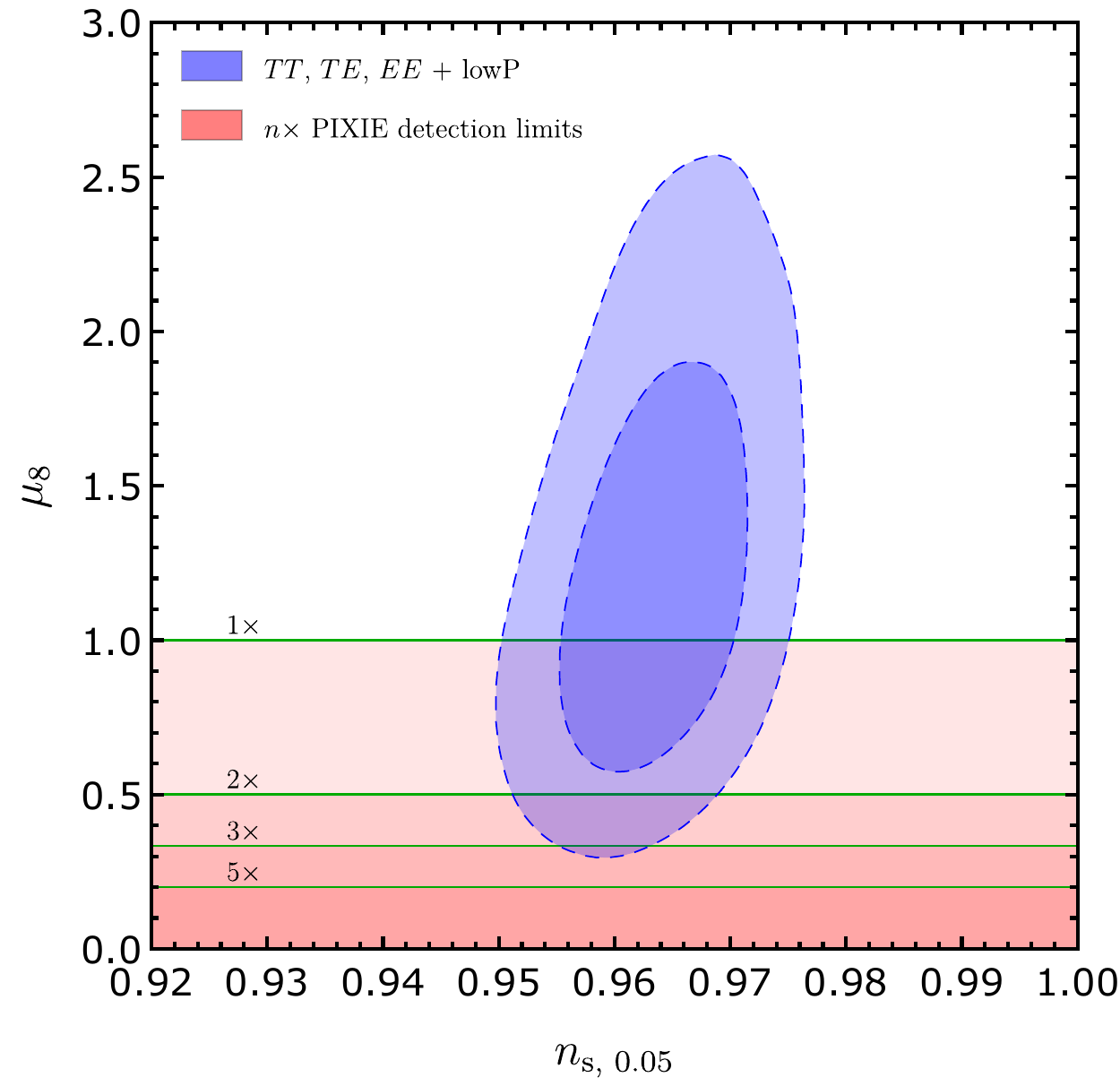}
&\includegraphics[width=\columnwidth]{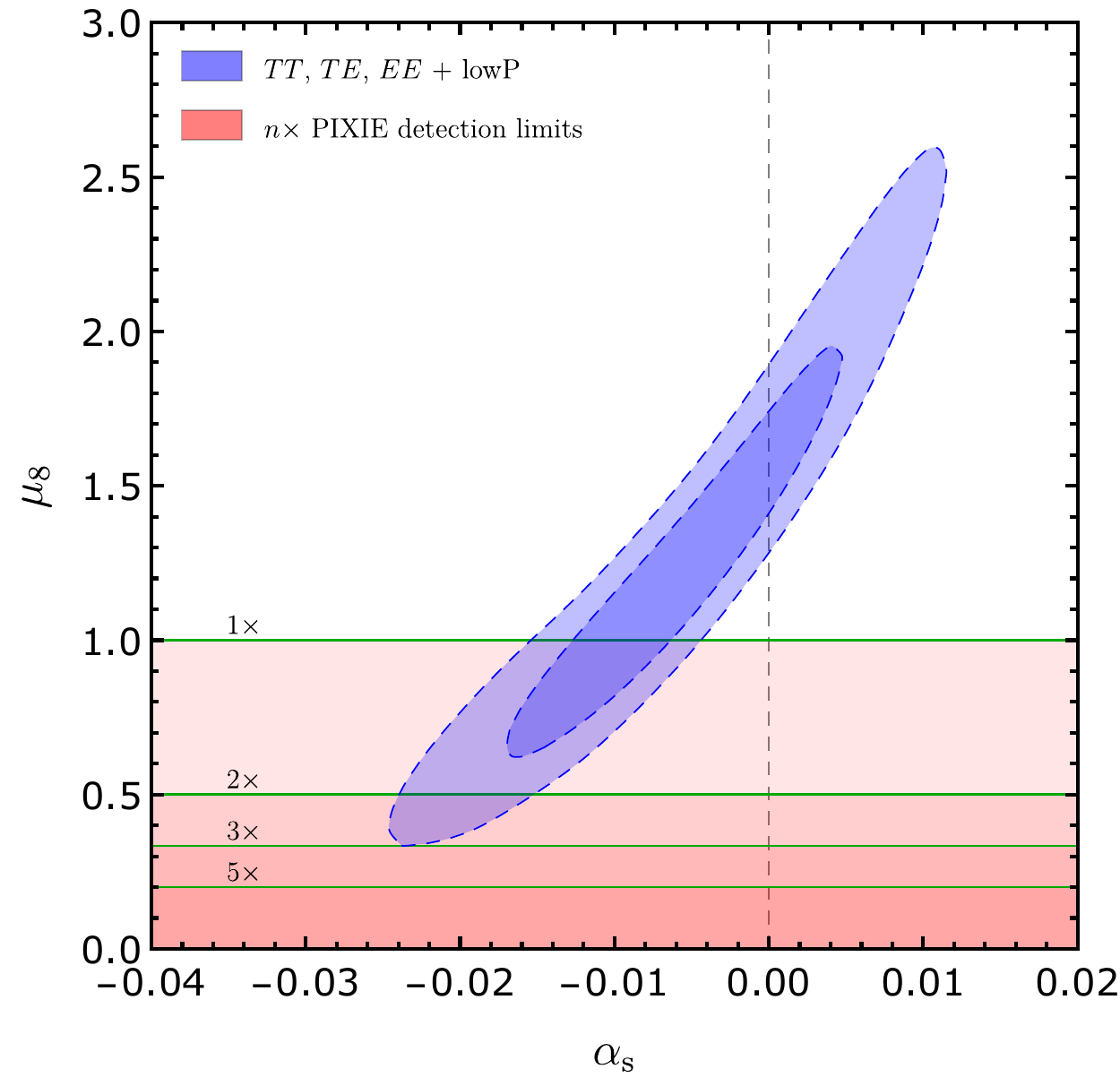}
\end{tabular}
\end{center}
\caption{\footnotesize{This figure shows the $\limit{68}$ and $\limit{95}$ contours in the $n_\tu{s}$ - $\mu$ (left panel) and the $\alpha_\tu{s}$ - $\mu_8$ plane (right panel) for the \emph{Planck} $TT$, $TE$, $EE$ + lowP dataset for $\Lambda\mathrm{CDM} +\nrun$, together with the $1\sigma$ detection limits for PIXIE and possible improvements.}}
\label{fig:ns-nrun-detectionlimits}
\end{figure*}

\noindent As we discussed in the previous section, the expected primordial spectral distortion $\mu$ is 
a function of cosmological parameters that play a role during the early universe epoch (like the scalar
spectral index $\ns$, its running $\nrun$, the cold dark matter energy density, etc.). 
Since most of these parameters are now well constrained by the recent measurements of
CMB anisotropies (both in temperature an polarization) made by the \emph{Planck} satellite,
one could, albeit indirectly, constrain the expected value of $\mu$ assuming 
a $\Lambda$CDM model or one of its extensions (see also \cite{Enqvist:2015njy} for a recent analysis). 

Spectral distortions in the $\mu$-
era can be computed in terms of $6$ - $7$ parameters (which we will call 
$\vec{\theta}$):
\begin{itemize}[leftmargin=*]
\item the baryon and cold dark matter density parameters $\Omega_\mathrm{b} h^2\equiv\omega_\mathrm{b}$ and $\Omega_\mathrm{c} h^2\equiv\omega_\mathrm{c}$, together with the number of effective relativistic degrees of freedom $N_\mathrm{eff}$. These enter in the computation of the expansion history: from them we compute the Hubble constant $H_0$ and 
the Helium mass fraction $Y_P$ that enter in the computation of the dissipation scale $k_\tu{D}$;
\item the CMB temperature $T_\mathrm{CMB}$;
\item the parameters describing the primordial spectrum,
\begin{equation}
\label{eq:pofk}
P_{\zeta}(k) = \frac{2\pi^2}{k^3}\Delta_\zeta(k) = A_\mathrm{s}\,\bigg(\frac{k}{k_\star}\bigg)^{n_\tu{s}-1 + \frac{\alpha_\tu{s}}{2}\log\frac{k}{k_\star}}\,,
\end{equation}
namely the amplitude $\log(\num{d10} A_\mathrm{s})$ and tilt $n_\mathrm{s}$ for the $\Lambda\text{CDM}$ case, with the addition of the running $\nrun$ for the $\Lambda\text{CDM} + \nrun$ case.
\end{itemize}

We performed an analysis of the recent \emph{Planck} $TT$, $TE$, $EE$ + lowP likelihood \cite{Aghanim:2015xee}, which includes the 
(temperature and $E$-mode polarization) high-$\ell$ likelihood together with the $TQU$ pixel-based low-$\ell$ likelihood, through Monte Carlo Markov Chain sampling, using the publicly available code \texttt{cosmomc}~\cite{Lewis:2002ah, Lewis:2013hha}. 
We have varied the primordial parameters, along with $\omega_\mathrm{b}$, $\omega_\mathrm{c}$, 
the reionization optical depth $\tau$, and finally the ratio of the sound horizon to the 
angular diameter distance at decoupling $100\,\theta_\mathrm{MC}$. For each model in the MCMC chain we compute, 
as derived parameter, the value of $\mu_8$ using the ``\texttt{idistort}'' code developed by Khatri and Sunyaev \cite{Khatri:2012tw, Khatri:2013dha}. 
For this purpose we fix the CMB temperature to $T_\mathrm{CMB} = 2.7255\,\mathrm{K}$, the neutrino 
effective number to the standard value $N_\mathrm{eff}=3.046$, and we evaluate the 
primordial Helium abundance $Y_P$ assuming standard Big Bang Nucleosynthesis. 

Processing the chains through the \texttt{getdist} routine (included in the \texttt{cosmomc}~package), and marginalizing over all the nuisance parameters, we obtain for the $\Lambda$CDM case (no running) the indirect constraint 
$\mu_8 = 1.57^{+0.108}_{-0.127}$ at $\limit{68}$. 
Including the possibility of a running, the \emph{Planck} constraint on $\mu$ is weakened to $\mu_8 = 1.28^{+0.299}_{-0.524}$ 
($\limit{68}$). The marginalized posterior distributions for $\mu_8$ are shown in \fig{mu-1d}. Notice that the ``the balanced injection scenario'', namely the possibility that the negative contribution to $\mu$ from adiabatic cooling cancels precisely the positive contribution from the dissipation of adiabatic modes \cite{Khatri:2011aj,Chluba:2012gq}, leaving $\mu_8 = 0$, is excluded at extremely high significance (\ie $\approx 15\sigma$) for the $\Lambda\mathrm{CDM}$ model, and at $\limit{\num{97.4}}$\footnote{We quote the confidence level, in this case, because the 
posterior for $\mu_8$ is non-Gaussian (as can be seen from \fig{mu-1d}).} if we allow the running to vary.

\fig{ns-nrun-detectionlimits} shows the dependence of $\mu$-distortion on the tilt $n_\mathrm{s}$ and the running $\alpha_\tu{s}$:
\begin{itemize}[leftmargin=*]
\item in the left panel we see that $\mu_8$ is not very degenerate with $n_\mathrm{s}$. The reason is twofold. First and most importantly, for non-zero running of order $\num{d-2}$, as allowed by \emph{Planck}, a change in the tilt of order $\num{d-2}$ is a small correction to the power spectrum at the short scales that are responsible for $\mu$-distortions ($\nrun$ appears in \eq{pofk} with a factor of $(\log k/k_\star)/2\sim 5$ for $k\sim\num{d3}\,\mathrm{Mpc}^{-1}$). Secondly, \emph{Planck} constraints on $n_\mathrm{s}$ are tighter than those on $\nrun$ by roughly a factor of two;
\item the right panel, on the other hand, shows that $\mu_8$ is strongly dependent on $\nrun$ (increasing $\nrun$ increases the power at short scales and hence leads to a larger $\mu_8$, and viceversa). We also note that the two dimensional contour in the $\nrun$ - $\mu_8$ plane is not ellipsoidal, but banana-shaped. The reason is that at large negative running, the contribution to spectral distortions from dissipation of acoustic waves will go to zero asymptotically, and the net $\mu$ amplitude will be the one from adiabatic cooling (which for the tightly constrained values of cosmological parameters can be practically considered a constant).
\end{itemize}
Having discussed the current (indirect) limits on $\mu$-distortions from \emph{Planck} measurements of CMB temperature and polarization anisotropies, we move to the forecasts for a PIXIE-like spectrometer.


\section{Forecasts for PIXIE: Fisher analysis}
\label{sec:fisher}

\begin{figure*}
\begin{center}
\begin{tabular}{c c}
\includegraphics[width=\columnwidth]{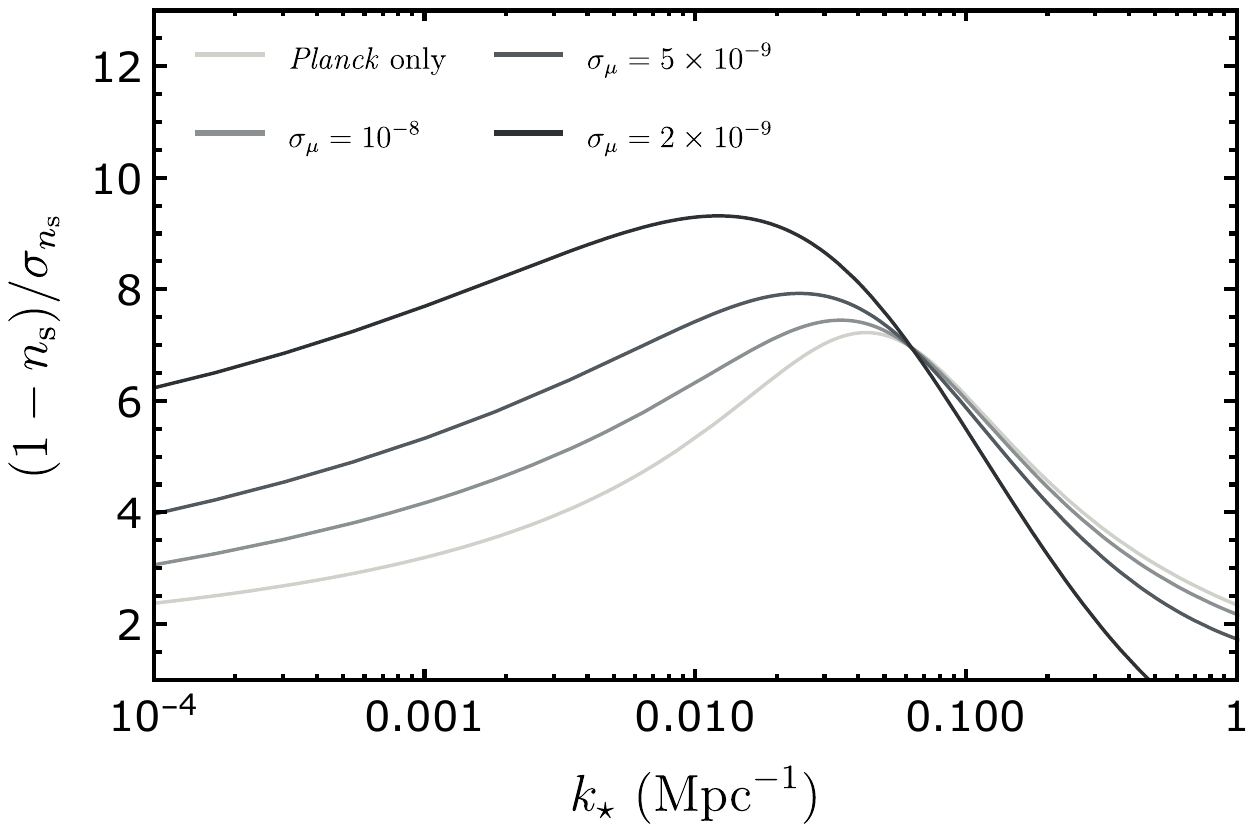}
&\includegraphics[width=0.9837\columnwidth]{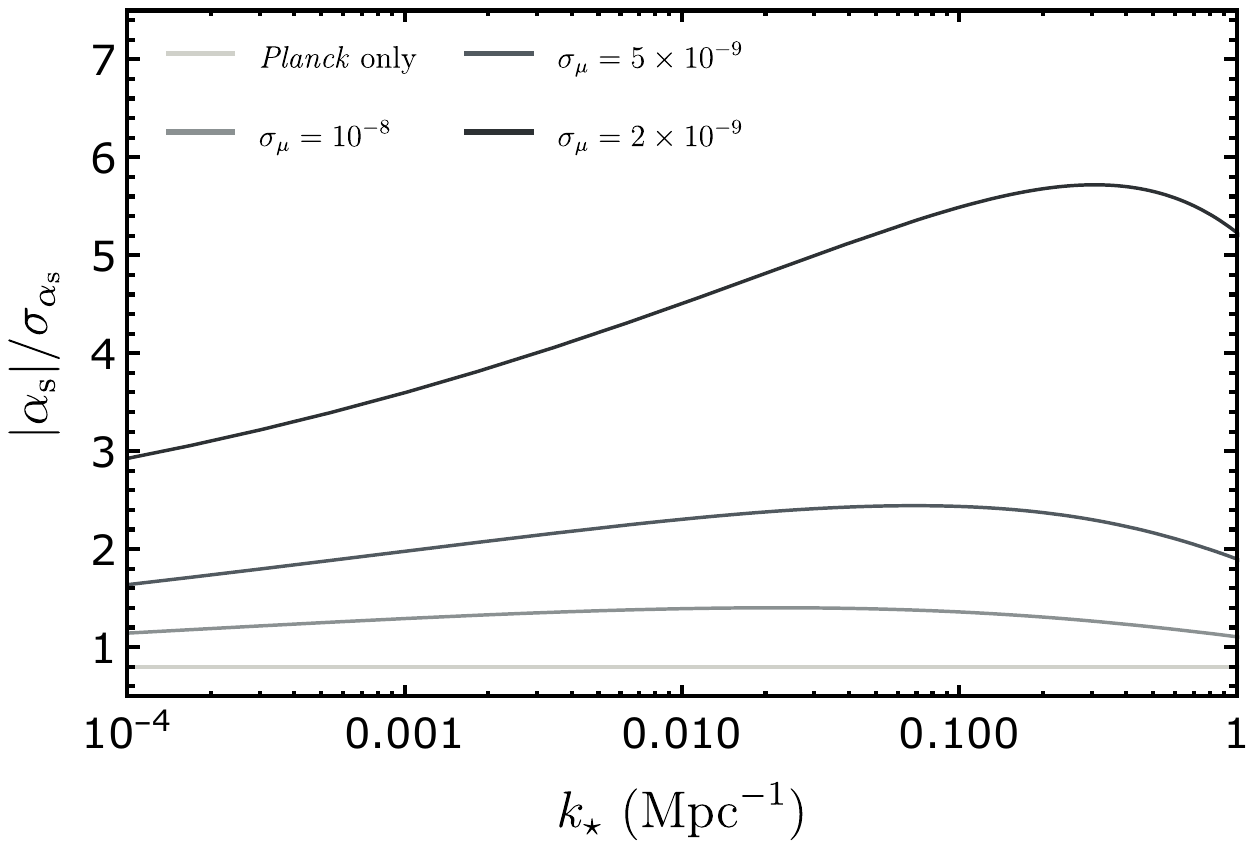} 
\end{tabular}
\end{center}
\caption{\footnotesize{$(1 - n_\tu{s})/\sigma_{n_\tu{s}}$ (left panel) and $\abs{\alpha_\tu{s}}/\sigma_{\alpha_\tu{s}}$ (right panel) as function of the pivot scale $k_\star$, for a vanishing fiducial distortions $\mufid_8 = 0$. A dependence on the pivot scale is always present for $\ns$ (left panel), while for $\nrun$ the dependence becomes appreciable only for a significant improvements over PIXIE's sensitivity. The optimal choice of $k_\star$ shifts towards $k < 0.05\,\mathrm{Mpc}^{-1}$ for $n_\tu{s}$ and $k > 0.05\,\mathrm{Mpc}^{-1}$ for $\nrun$, when the information from spectral distortions is included.}}
\label{fig:detection}
\end{figure*}

\begin{figure*}
\begin{center}
\begin{tabular}{c c}
\includegraphics[width=\columnwidth]{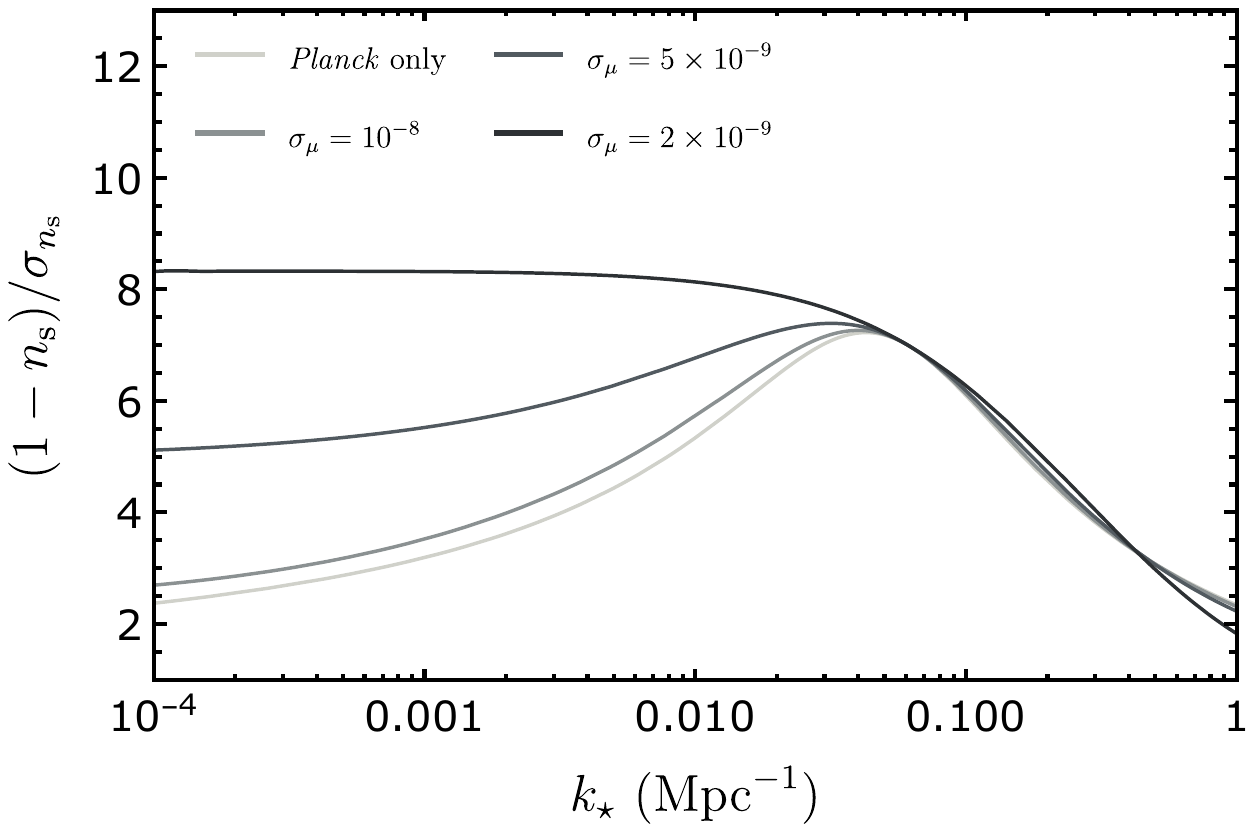}
&\includegraphics[width=1.015\columnwidth]{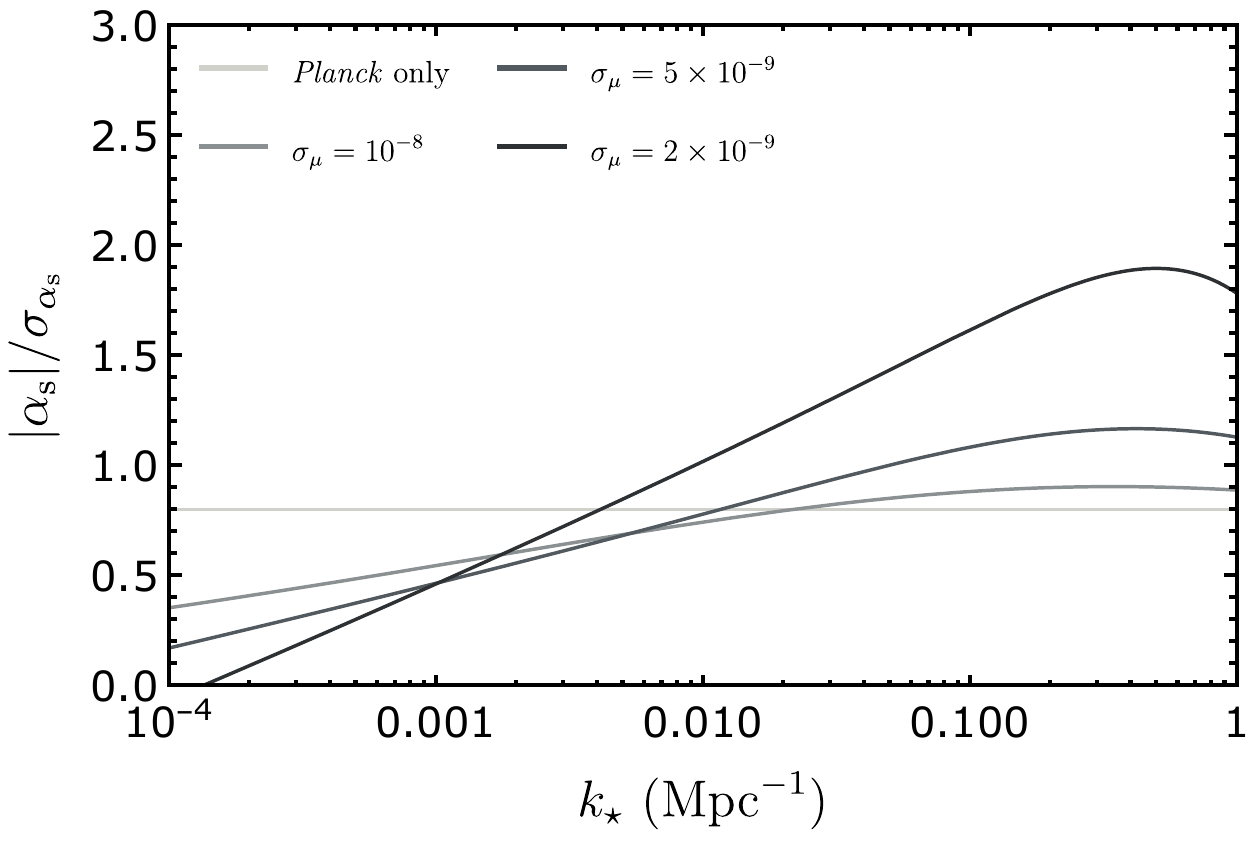}
\end{tabular}
\end{center}
\caption{\footnotesize{Same as \fig{detection}, but in this case we consider a fiducial $\mu_8$ amplitude of $\num{1.3}$ 
(see \sect{fisher} for details). The behavior is qualitatively similar to the case of a vanishing fiducial $\mu_8$.}}
\label{fig:detection-fid_neq_0}
\end{figure*}

\noindent {Considering only $\mu$-distortions, and using the approximation in terms of a window function from $z_1 = \num{5d4}$ to $z_2 = \num{2d6}$ (with the amplitude of the scalar spectrum fixed at $A_\mathrm{s}\approx\num{2.2d-9}$), we can perform a simple Fisher forecast to see how the constraints on tilt and running are improved by combining PIXIE with the \emph{Planck} prediction for $\mu_8$.}

{This allows us also to discuss, mirroring what has been done for CMB anisotropies alone in }\cite{Cortes:2007ak}{, what is the optimal choice of pivot scale (that maximizes the detection power for these two parameters) for the combined analysis, as function of the sensitivity of a PIXIE-like mission. We stress that the choice of pivot has no impact on the detectability of $\mu$-distortions themselves: it is just a particular way to parametrize the spectrum. Whether or not $\mu$-distortions will be seen is only dependent on the amount of scalar power at small scales (which is captured by the fiducial $\mu_8$ that we consider).}

{Finally, we point out that an improvement of a factor of three over PIXIE implies a guaranteed discovery.}

We add to the \emph{Planck} bounds the detection limits for $\mu$-distortions from the PIXIE white paper \cite{Kogut:2011xw}, \ie 
\begin{equation}
\label{eq:fisher-joint-like}
\begin{split}
\mathcal{L}(n_\tu{s}, \alpha_\tu{s})&\propto\mathcal{L}(n_\tu{s}, \alpha_\tu{s})_\tu{\emph{Planck}}\,\times \\
&\;\;\;\;\exp\bigg[-\frac{(\mu_8(n_\tu{s}, \alpha_\tu{s}) + \mu_{8,\mathrm{BEC}} - \mufid_8)^2}{2\sigma_{\mu_8}^2}\bigg]\,\,,
\end{split}
\end{equation}
with $\sigma_{\mu_8} = \num{1}$ ($\num{0.5}$ and $\num{0.2}$) for ($2\times$ and $5\times$) PIXIE, 
and $\mu_8(n_\tu{s}, \alpha_\tu{s})$ given by
\begin{equation}
\label{eq:noncera}
\begin{split}
\mu_8(n_\tu{s}, \alpha_\tu{s}) &= \num{2.3d8}\times \\
&\;\;\;\;A_\mathrm{s}\int_{k_\tu{D}(z_1)}^{k_\tu{D}(z_2)}\frac{\dif k}{k}\bigg(\frac{k}{k_\star}\bigg)^{n_\tu{s}-1 + \frac{\alpha_\tu{s}}{2}\log\frac{k}{k_\star}}\,\,.
\end{split}
\end{equation}
The results of this Fisher analysis are just an approximation of the full MCMC sampling of the joint likelihood that we will present in the next section. 
We can then safely consider only two fiducial values for $\mu_8$, which approximate well the choices we will make later (see \tab{future}): 
\begin{itemize}[leftmargin=*]
\item $\mufid_8 = 0$, \ie a cosmology with zero $\mu$-type distortions;
\item $\mufid_8 = 1.3$, \ie the mean-fit value from \emph{Planck} data (for the $\Lambda\text{CDM} + \nrun$ case).
\end{itemize}

As \emph{Planck} likelihood, we take (disregarding for simplicity the normalization)
\begin{equation}
\label{eq:fisher-planck-like}
\begin{split}
\log\mathcal{L}(n_\tu{s}, \alpha_\tu{s})_\tu{\emph{Planck}} &= -\frac{(\alpha_\tu{s} - \bar{\alpha}_\tu{s})^2}{2\sigma^2_{\alpha_\tu{s}}} \\
&\;\;\;\;- \frac{(n_\tu{s} + \frac{\alpha_\tu{s}}{2}\log (k_\star/k_\star^{(0)}) - \bar{n}_\tu{s})^2}{2\sigma^2_{n_\tu{s}}}\,\,,
\end{split}
\end{equation}
where:
\begin{itemize}[leftmargin=*]
\item the tilt is written at an arbitrary pivot $k_\star$ in terms of the running and the reference scale $k_\star^{(0)}$ (note that the Jacobian of the transformation is $1$ so it can be neglected); 
\item $k_\star^{(0)} = 0.05\,\mathrm{Mpc}^{-1}$ is the scale where $n_\tu{s}$ and $\alpha_\tu{s}$ decorrelate: for this reason we take $\bar{n}_\tu{s}$, $\bar{\alpha}_\tu{s}$ to be the marginalized means from the \emph{Planck} $TT$, $TE$, $EE$ + lowP analysis. $\sigma_{n_\tu{s}}$ and $\sigma_{\alpha_\tu{s}}$ are the corresponding marginalized standard deviations. The values are listed in \tab{fisher-mean+sigmas}.
\end{itemize}

\begin{table}[!hbt]
\begin{center}
\begin{tabular}{cc}
\toprule
\horsp
$\bar{n}_\tu{s}$ \vertsp $0.963
9$ \\
\horsp
$\bar{\alpha}_\tu{s}$ \vertsp $\num{-0.0057}$ \\
\hline
\horsp
$\sigma_{n_\tu{s}}$ \vertsp 
$\num{0.0050}$ \\
\horsp
$\sigma_{\alpha_\tu{s}}$ \vertsp $\num{0.0071}$ \\
\botrule
\end{tabular}
\caption{\footnotesize{Mean and standard deviation for spectral index and running used in \eq{fisher-planck-like}, from the \emph{Planck} $TT$, $TE$, $EE$ + lowP analysis.}}
\label{tab:fisher-mean+sigmas}
\end{center}
\end{table}

\fig{detection} shows $(1 - n_\tu{s})/\sigma_{n_\tu{s}}$ and $\abs{\alpha_\tu{s}}/\sigma_{\alpha_\tu{s}}$ as function of the pivot scale for vanishing $\mufid$: we see that, as we increase the sensitivity of PIXIE, the $k_\star$ that maximizes the detection of the tilt is shifted towards values smaller than $k_\star^{(0)} = \num{0.05}\,\mathrm{Mpc}^{-1}$. The best pivot for the running moves in the opposite direction, towards values larger than $
\num{0.05}\,\mathrm{Mpc}^{-1}$. \fig{detection-fid_neq_0} shows that the same qualitative behavior is reproduced in the case of a fiducial $\mu_8$ different from zero.

These plots show that the effect of changing of pivot on the detection power for $\ns$ and $\nrun$ is not very relevant, if we increase {$1/\sigma_{\mu_8}$} up to $5\times$ PIXIE. At $10\times$ the choice of $k_\star$ can lead to a small improvement on $\sigma_{\nrun}$: this is an interesting result, that could open up the possibility of choosing the pivot 
outside of the CMB window in the future, as $\sigma_{\mu_8}$ becomes even smaller.\footnote{However, from the left panels of Figs.~\ref{fig:detection} and \ref{fig:detection-fid_neq_0} we see how this improvement would be at the expense of an increased error on the tilt $\ns$.} However, since we will stop at $10\times$ PIXIE (\ie the expected error on $\mu_8$ achievable by PRISM) in this work, we will keep $k_\star = \num{0.05}\,\mathrm{Mpc}^{-1}$ in the 
following sections.

For vanishing $\mufid_8$, \fig{detection} shows that the improvement for $\sigma_{\nrun}$ can be greater than the case with non-zero fiducial. However, it is important to stress that the assumption of having zero distortions starts to become incompatible with the \emph{Planck} indirect constraints on $\mu_8$ (as one can see, \textit{e.g.}, from \fig{mu-1d}) for $\sigma_{\mu_8}\approx\num{0.3}$, making a combination of the two likelihoods inadvisable (this is also the reason why we have decided to not consider, in \sect{introduction}, a fiducial running so small that spectral distortions from Silk damping are absent). For 
fiducial $\mu_8$ different from zero we see that this does not happen: the combination of the likelihoods, which we will explore through MCMC sampling in the next section, is therefore justified in this case.

Finally, it is interesting to ask whether there exist any threshold value for sensitivity to 
the $\mu$ amplitude such that, by reaching it, we are guaranteed to learn something about the early universe, irrespectively of what the running might actually be. The right panel of \fig{detection} suggests the answer to this question (which we will confirm in the next section with a detailed calculation). Within the uncertainty of \emph{Planck}, a vanishing running implies a distortion of order $\mu_{8}\sim 1.4$, as we have seen in \sect{expectations}: therefore a measurement of the CMB spectrum at sensitivity of $\sigma_{\mu_{8}}\sim 1.4/4= 0.35$, corresponding to about $3\times$ PIXIE, must lead to\footnote{This assumes that we interpret the data within $\Lambda$CDM plus running. Given our theoretical understanding of the early universe, this is indeed perhaps the most natural choice.} a 
first detection of $\mu$-distortions or a detection of negative running, or both. In fact, any central value $\mu_{8}\lesssim 1.4/2\sim0.7$ at this resolution would exclude $\nrun\geq0$, while any larger $\mu_{8}$ would exclude $\mu_{8}\leq 0$ at $\limit{95}$.


\section{Forecasts for PIXIE: MCMC}
\label{sec:mcmc}

\begin{table*}[!hbtp]
\begin{center}
\begin{tabular}{cccc}
\toprule
\horsp
$TT$, $TE$, $EE$ + lowP \vertsp $n_\mathrm{s}$ \vertsp $\alpha_\mathrm{s}$ \vertsp $\mu_8$ \\
\hline
\horsp
$\Lambda\text{CDM}$ \vertsp $0.9645^{+0.0048}_{-0.0049}$ \vertsp $\equiv 0$ \vertsp $1.57^{+0.1
1}_{-0.1
3}$ \\
\horsp
$\Lambda\text{CDM} + \nrun$ \vertsp ${\num[round-mode=places,round-precision=4]{0.963863}}\pm0.0050$ \vertsp ${\num[round-mode=places,round-precision=4]{-0.00568968}}^{+0.0071}_{-0.0070}$ \vertsp ${\num[round-mode=places,round-precision=2]{1.27577}}^{+\num[round-mode=places,round-precision=2]{0.29864}}_{-\num[round-mode=places,round-precision=2]{0.524326}}$ \\
\horsp
``slow-roll'' \vertsp 
$0.9644^{+0.0051}_{-0.0052}$ \vertsp $\sim-(1-\ns)^2$ \vertsp 
$1.49_{-0.13}^{+0.12}$ \\
\hline
\hline
\horsp
$\nrunfid = -0.01$ ($\mufid_8 = 1.06$) \vertsp $n_\mathrm{s}$ \vertsp $\alpha_\mathrm{s}$ \vertsp $\mu_8$ \\
\hline
\horsp
\emph{Planck} + $1\times$ PIXIE \vertsp ${\num[round-mode=places,round-precision=4]{0.963682}}^{+\num[round-mode=places,round-precision=4]{0.0049693}}_{-\num[round-mode=places,round-precision=4]{0.0049002}}$ \vertsp ${\num[round-mode=places,round-precision=4]{-0.0063928}}^{+\num[round-mode=places,round-precision=4]{0.0065104}}_{-\num[round-mode=places,round-precision=4]{0.00642522}}$ \vertsp ${\num[round-mode=places,round-precision=2]{1.22102}}^{+\num[round-mode=places,round-precision=2]{0.283535}}_{-\num[round-mode=places,round-precision=2]{0.446506}}$ \\
\horsp
\emph{Planck} + $2\times$ PIXIE \vertsp ${\num[round-mode=places,round-precision=4]{0.963399}}^{+\num[round-mode=places,round-precision=4]{0.0048958}}_{-\num[round-mode=places,round-precision=4]{0.0048254}}$ \vertsp ${\num[round-mode=places,round-precision=4]{-0.00736549}}^{+\num[round-mode=places,round-precision=4]{0.00612838}}_{-\num[round-mode=places,round-precision=4]{0.00527112}}$ \vertsp ${\num[round-mode=places,round-precision=2]{1.15008}}^{+\num[round-mode=places,round-precision=2]{0.25372}}_{-\num[round-mode=places,round-precision=2]{0.344389}}$ \\
\horsp
\emph{Planck} + $3\times$ PIXIE 
\vertsp 
${\num[round-mode=places,round-precision=4]{0.963235}}
\pm0.0048$
\vertsp
${\num[round-mode=places,round-precision=4]{-0.00793214}}^{+\num[round-mode=places,round-precision=4]{0.00531736}}_{-\num[round-mode=places,round-precision=4]{0.00447949}}$
\vertsp
${\num[round-mode=places,round-precision=2]{1.10869}}^{+\num[round-mode=places,round-precision=2]{0.222486}}_{-\num[round-mode=places,round-precision=2]{0.270943}}$
\\
\horsp
\cellcolor{lightgray} \emph{Planck} + $5\times$ PIXIE \vertsp ${\num[round-mode=places,round-precision=4]{0.96313}}^{+\num[round-mode=places,round-precision=4]{0.0047631}}_{-\num[round-mode=places,round-precision=4]{0.004713}}$ \vertsp \cellcolor{lightgray} ${\num[round-mode=places,round-precision=4]{-0.00834119}}^{+\num[round-mode=places,round-precision=4]{0.00400554}}_{-\num[round-mode=places,round-precision=4]{0.00351656}}$ \vertsp ${\num[round-mode=places,round-precision=2]{1.07565}}^{+\num[round-mode=places,round-precision=2]{0.16992}}_{-\num[round-mode=places,round-precision=2]{0.18412}}$ \\
\horsp
\emph{Planck} + $10\times$ PIXIE \vertsp ${\num[round-mode=places,round-precision=4]{0.963149}}
\pm0.0047$ 
\vertsp ${\num[round-mode=places,round-precision=4]{-0.00845063}}^{+\num[round-mode=places,round-precision=4]{0.0025015}}_{-\num[round-mode=places,round-precision=4]{0.0024441}}$ \vertsp ${\num[round-mode=places,round-precision=2]{1.06181}}
\pm0.09$ \\
\hline
\hline
\horsp
$\nrunfid = -0.02$ ($\mufid_8 = 0.73$) \vertsp $n_\mathrm{s}$ \vertsp $\alpha_\mathrm{s}$ \vertsp $\mu_8$ \\
\hline
\horsp
\emph{Planck} + $1\times$ PIXIE \vertsp ${\num[round-mode=places,round-precision=4]{0.963468}}^{+\num[round-mode=places,round-precision=4]{0.0049678}}_{-\num[round-mode=places,round-precision=4]{0.0049065}}$ \vertsp ${\num[round-mode=places,round-precision=4]{-0.00714359}}^{+\num[round-mode=places,round-precision=4]{0.00647366}}_{-\num[round-mode=places,round-precision=4]{0.00632084}}$ \vertsp ${\num[round-mode=places,round-precision=2]{1.17611}}^{+\num[round-mode=places,round-precision=2]{0.272146}}_{-\num[round-mode=places,round-precision=2]{0.426209}}$ \\
\horsp
\emph{Planck} + $2\times$ PIXIE \vertsp ${\num[round-mode=places,round-precision=4]{0.962833}}^{+\num[round-mode=places,round-precision=4]{0.0049068}}_{-\num[round-mode=places,round-precision=4]{0.0048354}}$ \vertsp ${\num[round-mode=places,round-precision=4]{-0.00936563}}^{+\num[round-mode=places,round-precision=4]{0.00614803}}_{-\num[round-mode=places,round-precision=4]{0.00523659}}$ \vertsp ${\num[round-mode=places,round-precision=2]{1.04236}}^{+\num[round-mode=places,round-precision=2]{0.230696}}_{-\num[round-mode=places,round-precision=2]{0.313091}}$ \\
\horsp
\cellcolor{lightgray} \emph{Planck} + $3\times$ PIXIE \vertsp
${\num[round-mode=places,round-precision=4]{0.962356}}^{+\num[round-mode=places,round-precision=4]{0.0049068}}_{-\num[round-mode=places,round-precision=4]{0.0048192}}$
\vertsp
\cellcolor{lightgray} ${\num[round-mode=places,round-precision=4]{-0.0110635}}^{+\num[round-mode=places,round-precision=4]{0.00548617}}_{-\num[round-mode=places,round-precision=4]{0.00454196}}$
\vertsp
${\num[round-mode=places,round-precision=2]{0.951016}}^{+\num[round-mode=places,round-precision=2]{0.194843}}_{-\num[round-mode=places,round-precision=2]{0.242651}}$
\\
\horsp
\emph{Planck} + $5\times$ PIXIE \vertsp ${\num[round-mode=places,round-precision=4]{0.961785}}^{+\num[round-mode=places,round-precision=4]{0.0048609}}_{-\num[round-mode=places,round-precision=4]{0.0047376}}$ \vertsp ${\num[round-mode=places,round-precision=4]{-0.0130826}}^{+\num[round-mode=places,round-precision=4]{0.00461397}}_{-\num[round-mode=places,round-precision=4]{0.00367582}}$ \vertsp ${\num[round-mode=places,round-precision=2]{0.851889}}^{+\num[round-mode=places,round-precision=2]{0.155126}}_{-\num[round-mode=places,round-precision=2]{0.15369}}$ \\
\horsp
\emph{Planck} + $10\times$ PIXIE \vertsp ${\num[round-mode=places,round-precision=4]{0.961281}}^{+\num[round-mode=places,round-precision=4]{0.0047954}}_{-\num[round-mode=places,round-precision=4]{0.004716}}$ \vertsp ${\num[round-mode=places,round-precision=4]{-0.0148766}}^{+\num[round-mode=places,round-precision=4]{0.00329817}}_{-\num[round-mode=places,round-precision=4]{0.00291553}}$ \vertsp ${\num[round-mode=places,round-precision=2]{0.77074}}
\pm0.09$ \\
\botrule
\end{tabular}
\caption{\footnotesize{$\limit{68}$ constraints on the scalar spectral index $\ns$, its running $\nrun$ and the $\mu$-distortion amplitude from
a future combined analysis of the \emph{Planck} 2015 release in temperature and polarization and a PIXIE-like spectrometer as function of different experimental configurations and fiducial values for $\mu_8$. Notice that for $\nrunfid=-0.01$, $5\times$ PIXIE is needed to exclude $\nrun=0$ at $\limit{95}$, while for $\nrunfid=-0.02$, $3\times$ PIXIE suffices.}}
\label{tab:future}
\end{center}
\end{table*}

\begin{figure*}
\begin{center}
\begin{tabular}{c c}
\includegraphics[width=\columnwidth]{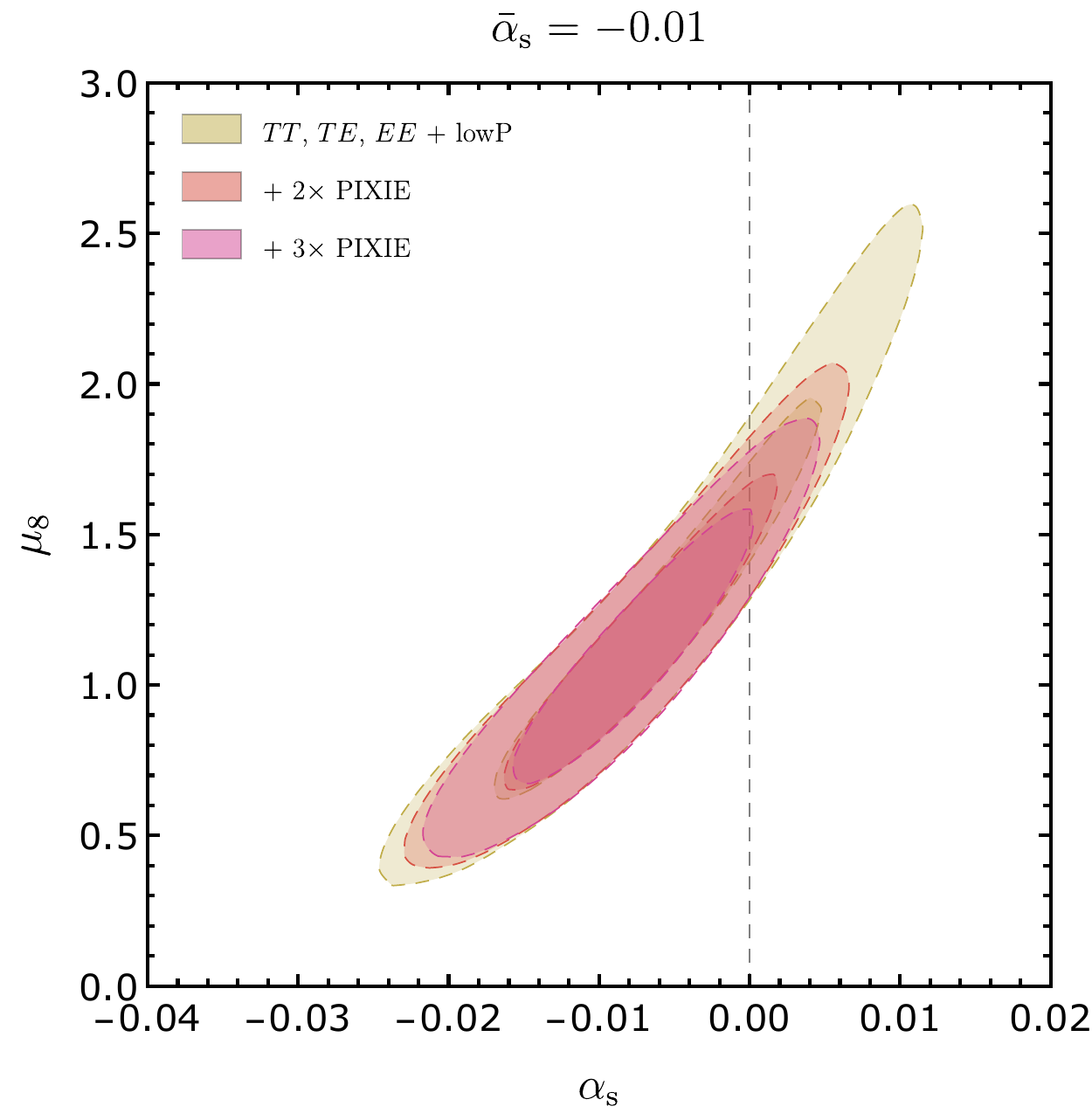}
&\includegraphics[width=\columnwidth]{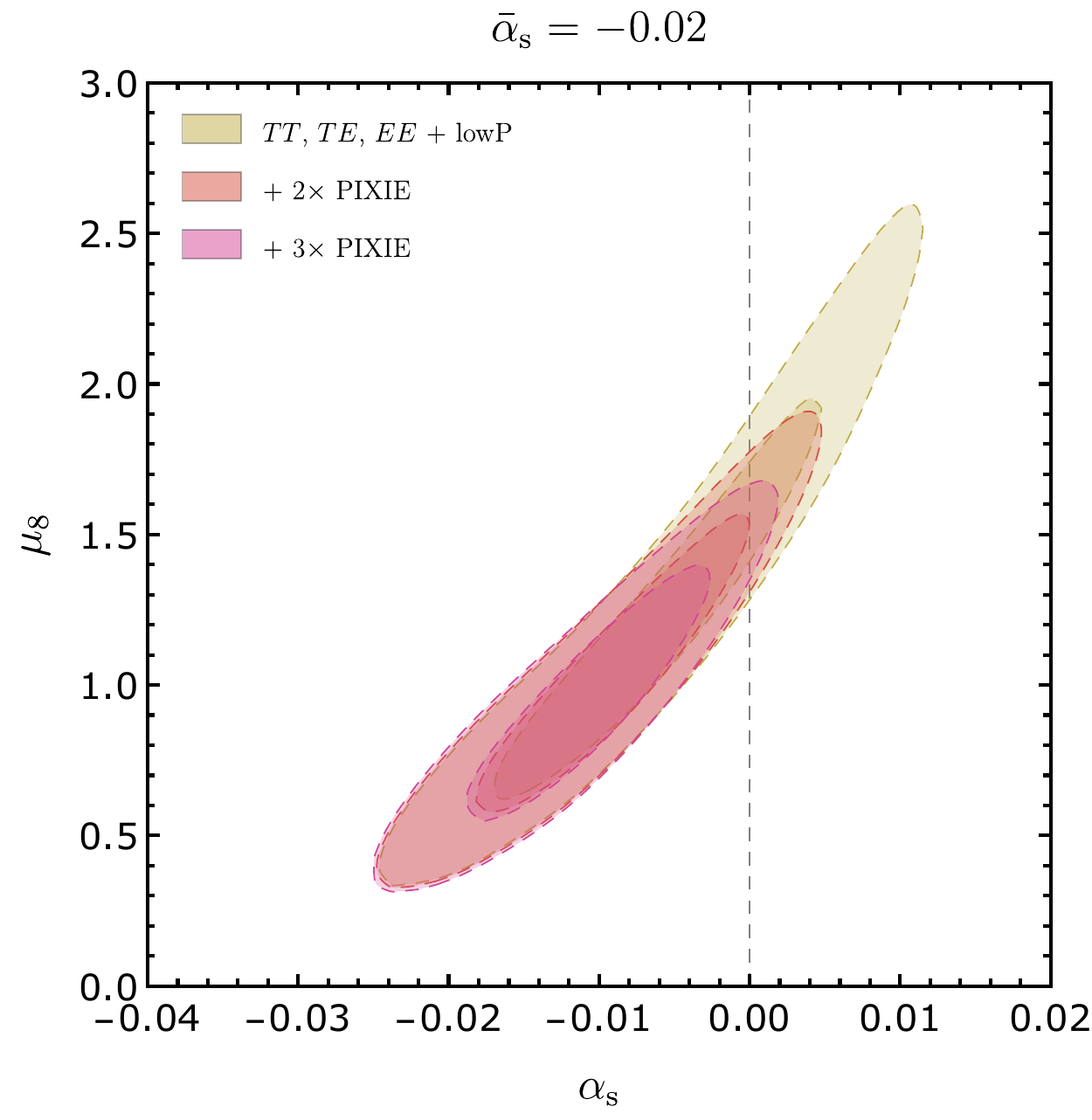} 
\end{tabular}
\end{center}
\caption{\footnotesize{Left panel: $\limit{68}$ and $\limit{95}$ contours in the $\nrun$ - $\mu_8$ plane, for \emph{Planck} alone (yellow contour) and including in the analysis the likelihood with $\nrunfid = -0.01$ (\ie $\mufid_8 = 1.06$) for a 
$2\times$ and $3\times$ 
improvement over PIXIE (orange and purple contours). Right panel: same as left panel, but with fiducial $\nrun$ equal to $-0.02$.}}
\label{fig:nrun-detectionlimits-forecast}
\end{figure*}

\noindent This section contains the main results of the paper, summarized in \tab{future} and 
\fig{nrun-detectionlimits-forecast} (which shows the contours in the $\nrun$ - $\mu_8$ plane). 

We start with a discussion of the detectability of $\mu$-type distortions by PIXIE in the context of the $\Lambda$CDM model, \ie with zero running of the spectral index. We stress that, in this case, \emph{Planck} bounds imply that with only a small $2\times$ improvement over PIXIE's noise, the exclusion of $\mu_8\leq 0$ at $\approx3\sigma$ is guaranteed, given the narrow posterior for $\mu_8$. 

On the other hand, as we have seen in \sect{expectations}, for the $\Lambda\mathrm{CDM} + \nrun$ case a value of $\mu_8\sim 0.7$ is fully compatible with \emph{Planck} data, and it will be only marginally detectable by PIXIE in the case of a minimal 
configuration. 
Assuming the \emph{Planck} constraint on $\mu_8$, the minimal value of $\mu_8$ compatible with \emph{Planck} in
between two standard deviations is $\mu_8 \sim 0.25$. Clearly, given this value, a safe 
experimental direct detection of $\mu$-type distortions can be obtained only with an experimental
sensitivity of $\sigma_{\mu_8} \sim 0.2$, \ie a $5\times$ improvement over PIXIE. 

However, in the presence of running, the argument can be reversed: it becomes now interesting to see how precise should be the
measurement performed by a PIXIE-like spectrometer to translate a non-detection of $\mu$-distortions into a detection of $\nrun < 0$, pursuing the marginal (below one standard 
deviation) indication for negative running coming from \emph{Planck} (whose 
posterior, while compatible with $\nrun = 0$, peaks at a negative value of 
$\nrun =-0.006$: see \tab{future}). 
For this purpose, we reprocess the MCMC chains by importance sampling, 
multiplying the weight of each sample by (see also \eq{fisher-joint-like})
\begin{equation}
\label{eq:post_process-PIXIE}
\mathcal{L}_\mathrm{PIXIE} = \exp\bigg[-\frac{(\mu_8(\vec{\theta}) + \mu_{8,\mathrm{BEC}}(\vec{\theta}) - \mufid_8)^2}{2\sigma^2_{\mu_8}}\bigg]\,\,,
\end{equation}
focussing on the two fiducial models for the running described in \sect{introduction}: 
\begin{itemize}[leftmargin=*]
\item $\nrunfid = -0.01$, 
corresponding to a spectral distortion $\mu_8 = 1.06$ (close to the \emph{Planck} best-fit for $\mu_8$);
\item $\nrunfid = -0.02$ corresponding to a spectral distortion $\mu_8 = 0.73$, which is at the limit of two standard deviations
from the \emph{Planck} mean-fit.
\end{itemize}
As in \sect{fisher}, we take $\sigma_{\mu_8} = 1/n$ for a $n\times$ PIXIE experimental configuration. The results of this importance sampling are also reported in \tab{future}.

Considering that from the \emph{Planck} dataset alone one obtains 
$\sigma_{\ns}\approx 0.005$ and $\sigma_{\nrun}\approx 0.007$, 
we see 
that the minimal configuration of $1\times$ PIXIE or the upgraded $2\times$ PIXIE 
will produce minimal effects on the \emph{Planck} bounds, even when the fiducial model deviates significantly from the \emph{Planck} best-fit. 
 
%

If, instead, the experimental sensitivity will reach the level of $5\times$ PIXIE ($10\times$ PIXIE) then the constraints 
on the running of the spectral index can be improved by $\sim 30\%$ ($\sim 50 \%$): this improvement could be extremely significant. More precisely, we see that if $\nrunfid$ is $\sim -0.01$, then
the addition of $5\times$ PIXIE to \emph{Planck} bounds could yield a detection of negative running at 
two standard deviations
(three for $10\times$ PIXIE). If we allow an even more negative fiducial value for the running, \ie $\nrunfid\sim -0.02$ then
the negative running will be probed at 
two standard deviations 
by 
$3\times$ PIXIE 
(five standard deviations 
by $10\times$ PIXIE
). \tab{future} also shows that the constraints on the tilt are left basically untouched, in agreement with the results of \sect{expectations}, where we have seen that $\mu_8$ is only mildly dependent on it.

Finally, we comment on the possibility of discriminating between no-running $\Lambda\text{CDM}$ and 
slow-roll inflation, where the running is 
second order in the slow-roll expansion. An order-of-magnitude prediction for $\nrun$, that arises in many models, is $\nrun\sim -(1-\ns)^2$ \cite{Roest:2013fha, Garcia-Bellido:2014gna, Gobbetti:2015cya}: \tab{future} shows that the predictions for $\mu_8$ in these two cases are indistinguishable at PIXIE's sensitivity, and that a massive improvement 
in sensitivity by a factor of order $\num{d
2}$ is needed to probe the differences between 
them. 


\section{Implications for slow-roll inflation}
\label{sec:slow-roll}

\noindent In this section, we discuss the implications of the value of the running within single-clock inflation. Observations tell us (see \cite{Ade:2015lrj} and \tab{future})

\bea
	1-\ns(k_{\star}) &\equiv & -\frac{\partial\log\Delta_\zeta^{2} (k_{\star})}{\partial\log k} \label{eq:tilt1} \\
	&=& \num{0.0361} \pm 0.0050 \quad \text{($\limit{68}$)}\,\,, \nonumber \\
	\nrun &=& -n_{\mathrm{s},N} \label{eq:tilt2} \\
	&=& -0.0057^{+0.0071}_{-0.0070} \quad \text{($\limit{68}$)}\,\,,\nonumber \\
	r&<&0.08 \quad \text{($\limit{95}$)}\,\,, \label{eq:tilt3}
\eea
where $\ast_{,N}$ refers to a derivative with respect to the number of e-foldings from the end of inflation, decreasing as time increases, namely $H\dif t=-\dif N$ (we refer to \cite{Baumann:2009ds} for a comprehensive review). The standard slow-roll solution for the primordial power spectrum gives (for an inflaton speed of sound $ c_\mathrm{s}$)
\bea
  1 - \ns &=& 2\epsilon-\frac{\epsilon_{,N}}{\epsilon}-\frac{c_{\mathrm{s},N}}{c_\mathrm{s}} \label{eq:eom1} \\
     &=& \frac{r}{8 c_\mathrm{s}} - \frac{r_{,N}}{r}\,\,, \label{eq:eom2} \\
     \nrun &=& 2\epsilon_{,N}-\frac{r_{,NN}}{r} + \bigg(\frac{r_{,N}}{r}\bigg)^{2}\,\,, \label{eq:eom3}
\eea
where the tensor-to-scalar ratio is given approximately by $r=16\epsilon c_\mathrm{s}$.

\begin{figure}[!hbt]
\includegraphics[width=.48\textwidth]{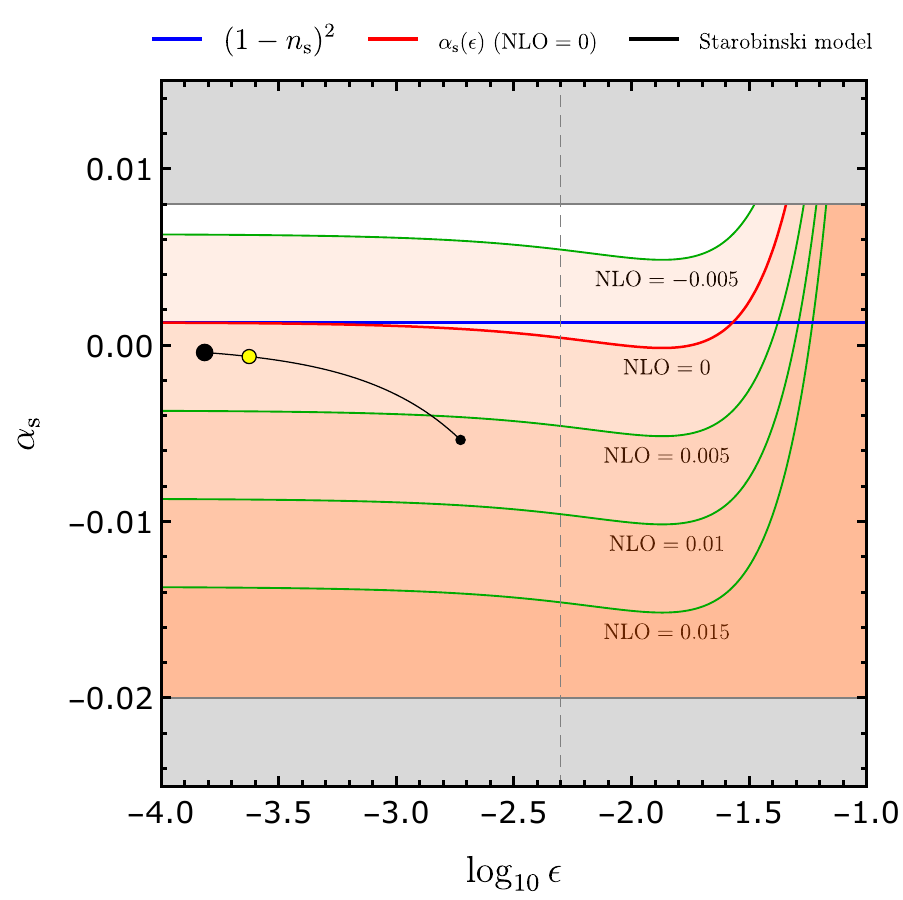}
\caption{\footnotesize{This contour plot shows $\nrun$ as function of 
$\epsilon$ for different values of the NLO slow-roll parameters. Notice that the uncertainty in $\ns$ is smaller than the thickness of the lines in the plot. In red we show $\alpha(\epsilon)$ of \eq{run-of-tilt} for $\mathrm{NLO} = 0$, while the blue line is its asymptotic value $(1-\ns)^{2}\approx 0.0013$. The black line 
shows the predictions of the Starobinsky model \cite{Starobinsky:1980te} (with $N$ going from $20$ to $70$), with the yellow dot being its prediction for $N = 56$ (chosen 
to reproduce the observed value of $\ns$). The gray bands show the values of $\nrun$ excluded (at $\limit{95}$) by \emph{Planck} $TT$, $TE$, $EE$ + lowP data, while the gray dashed vertical line shows the current bound on $\epsilon = r/(16c_\mathrm{s})$ from \eq{tilt3}, considering $c_\mathrm{s} = 1$.}}
\label{fig:NLOplot}
\end{figure}

It is convenient to re-express the running by making explicit its dependence on the tilt, which is relatively well constrained, \ie
\begin{equation}
\label{eq:run-of-tilt}
\nrun = (1-\ns)^{2} - 6\epsilon (1-\ns) + 8\epsilon^{2} - \bigg(\frac{rs}{8c_\mathrm{s}}+\frac{r_{,NN}}{r}\bigg)\,\,.
\end{equation}

Here, $\epsilon$ can be extracted from $r$ if we know the speed of sound $c_\mathrm{s}$ from the equilateral bispectrum, or if we assume standard slow-roll single-field inflation, namely $c_\mathrm{s}=1$. On the other hand, the last term $r_{,NN}/r$ makes its first appearance in the running $\nrun$; also the penultimate term $ s\equiv c_{\mathrm{s},N}/c_\mathrm{s} $ is degenerate with $ \epsilon_{,N}/\epsilon $ in $\ns$ and so it is also considered unknown. In this precise sense, we can think of the running as a measurement of the yet unknown next-to-leading (NLO) order slow-roll parameters
\be
\label{eq:NLO}
\mathrm{NLO}\equiv\frac{rs}{8c_\mathrm{s}} + \frac{r_{,NN}}{r} \quad \xrightarrow{c_\mathrm{s} = 1} \quad \frac{\epsilon_{,NN}}{\epsilon}\,\,.
\ee

In \fig{NLOplot}, we show a contour plot of $\nrun$ as function of $\epsilon$ for different values of the NLO slow-roll parameters. We point out that for $\text{NLO} = 0$ one finds {$\alpha \geq -\frac{1}{8}(1-\ns)^{2}\simeq -2\times 10^{-4} $}. Any evidence that the running is sizable and negative therefore implies $\text{NLO} > 0$, \ie the discovery of a new higher order slow-roll parameter. In a typical slow-roll model, one indeed expects the NLO terms to be of the same order as $(1-\ns)^{2}$. For example, consider $c_\mathrm{s} = 1$ and $\epsilon = 3/(4N^{2})$,\footnote{Note that the relation $\epsilon = 3/(4N^2)$ holds at first order in slow-roll: it is accurate enough, however, for the values of $N$ that reproduce a scalar spectral index $\ns$ within the current \emph{Planck} bounds.} \ie the Starobinsky model \cite{Starobinsky:1980te}. 
Then we have 
\bea
  \left( 1-\ns \right)^{2} &\simeq& \frac{4}{N^{2}}\,\,, \label{eq:starob1} \\
  \frac{r_{,NN}}{r} &=& \frac{6}{N^{2}}\,\,. \label{eq:starob2}
\eea
One hence finds 
\be
\label{eq:starob3}
\nrun\simeq -\frac{2}{N^{2}}\simeq -\frac{1}{2}(1-\ns)^{2}\,\,.
\ee


\section{Conclusion}
\label{sec:conclusion}

\noindent In this work, we have considered how a measurement of the CMB spectrum by an experiment like PIXIE 
would extend our knowledge of the very early universe. Using \emph{Planck} data, we have derived the predicted likelihood for the size of the $\mu$-type distortions generated by the dissipation of acoustic waves in the photon-baryon-electron plasma. As shown in \fig{mu-1d}, both $\Lambda$CDM and $\Lambda\mathrm{CDM} + \nrun$ predict $\mu_8\simeq\mathcal{O}(1)$, and exclude $\mu_8=0$, a.k.a. the ``the balanced injection scenario'' \cite{Chluba:2011hw,Khatri:2011aj,Chluba:2012gq} at high 
confidence (at $15\sigma$ for $\Lambda$CDM, at $\limit{\num{97.4}}$ for $\Lambda\mathrm{CDM} + \nrun$). 
While this means that we will be eventually able to measure $\mu$-distortions, it is important to determine whether this will already be possible with the next satellite experiment. Here we point out that, irrespectively of the actual value of $\nrun$ (and its respective $\mu_8$, according to \eq{noncera}), a meaningful sensitivity target is $\sigma_{\mu_8}\simeq 0.35$, namely about 
a three times improvement over the current PIXIE design (but still less sensitive than the proposed PRISM). This is in fact the threshold for a guaranteed discovery: either $\mu_8$ is large enough that it will be detected (at $\limit{95}$), or else $\nrun\geq 0$ will be excluded (at $\limit{95}$) and with it our current standard model, namely the $6$-parameter $\Lambda$CDM. The absence of a detection of $\mu_8$ 
for a $3\times$ PIXIE improvement would exclude most slow-roll models as well, since typically $\abs{\nrun} \sim (1-\ns)^{2}$, which is indistinguishable from $\nrun=0$ at these sensitivities.

We have further considered the constraining power of CMB spectral distortions combined with the current \emph{Planck} data. We have discussed how to optimize this analysis by choosing an appropriate pivot for the parameterization of the primordial power spectrum (see \fig{detection-fid_neq_0} and \fig{detection-fid_neq_0}). In \tab{future}, we present the improved constraints on the spectral tilt and its running from \emph{Planck} plus an $n$-fold improvement over PIXIE 
sensitivity. For a fiducial $\nrun=-0.01$, close to the fit for \emph{Planck}, one expects a detection of $\mu_8$ at $\limit{95}$ already with $2\times$ PIXIE. Conversely, for a fiducial $\nrun=-0.02$ (which is at the low $\limit{95}$ end of the \emph{Planck} constraint), $3\times$ PIXIE will already provide evidence (at 
$2\sigma$) of a sizable negative running. This would put pressure on the standard slow-roll paradigm, which leads to the typical expectation $\nrun\simeq (1-\ns)^{2}$ (see, \textit{e.g.}, \eq{run-of-tilt}). Finally, we proposed \fig{NLOplot} as a convenient and compact way to visualize the improving constraints on the tilt, running and tensor-to-scalar ratio.


\section*{Acknowledgments}

\noindent We would like to thank Eleonora Di Valentino for help in the MCMC analysis and importance sampling. We also thank Martina Gerbino and Massimiliano Lattanzi for useful comments on the latter. We are grateful to Raphael Flauger and Luca Pagano for useful discussions about forecasts for upcoming CMB experiments. \tblack{We would like to thank Jens Chluba for careful reading of the manuscript and useful comments.} G.C. and A.M. are supported by the research grant Theoretical Astroparticle Physics number 2012CPPYP7 under the program PRIN 2012 funded by MIUR and by TASP, iniziativa specifica INFN. E.P. is supported by the Delta-ITP consortium, a program of the Netherlands organization for scientific research (NWO) that is funded by the Dutch Ministry of Education, Culture and Science (OCW). G.C. would like to thank the Delta-ITP consortium and Utrecht University for the support and hospitality during two visits that 
have led to this work.


\section{Appendix}
\label{sec:appendix}

\subsection{$\mu$-distortion from energy release $E\to E + \delta E$}
\label{sec:appendix-release}

\noindent The relation $\mu\approx1.4\times\delta E/E$ can be understood with the following simple calculation, recalling that during the $\mu$-era the total number of photons is conserved. Taking a Bose-Einstein spectrum with energy $E + \delta E$ and small chemical potential $\mu$, and expressing the temperature $T$ in terms of energy and chemical potential as
\begin{equation}
\label{eq:app-T-v-mu+E}
T = \frac{\sqrt[4]{15}\sqrt[4]{E}}{\sqrt{\pi}}\bigg(1 + \frac{1}{4}\frac{\delta E}{E} + \frac{45\zeta(3)}{2\pi^4}\mu\bigg) + \mathcal{O}^2\,\,,
\end{equation}
one can find the relation between $\delta E$ and $\mu$ by requiring that the increase of energy is not accompanied by an increase in the number of photons, which remains equal to that of the original Planck spectrum, \ie 
\begin{equation}
\label{eq:app-BE-v-BB}
N_{\text{B-E}}(T(E+\delta E,\mu),\mu) = N_\mathrm{Planck}(E)\,\,.
\end{equation}
Solving for $\mu$, one finds \cite{Zeldovich:1969ff, Sunyaev:1970er} 
\begin{equation}
\label{eq:app-mu-of-deltaE}
\mu = \frac{9 \pi ^4\zeta (3)}{2(\pi ^6-405 \zeta (3)^2)}\frac{\delta E}{E} \approx 1.4\times\frac{\delta E}{E}\,\,.
\end{equation}

\subsection{Spectral shapes and $S/N$ for PIXIE}
\label{sec:appendix-shapes}

\begin{figure}[!hbt]
\includegraphics[width=0.48\textwidth]{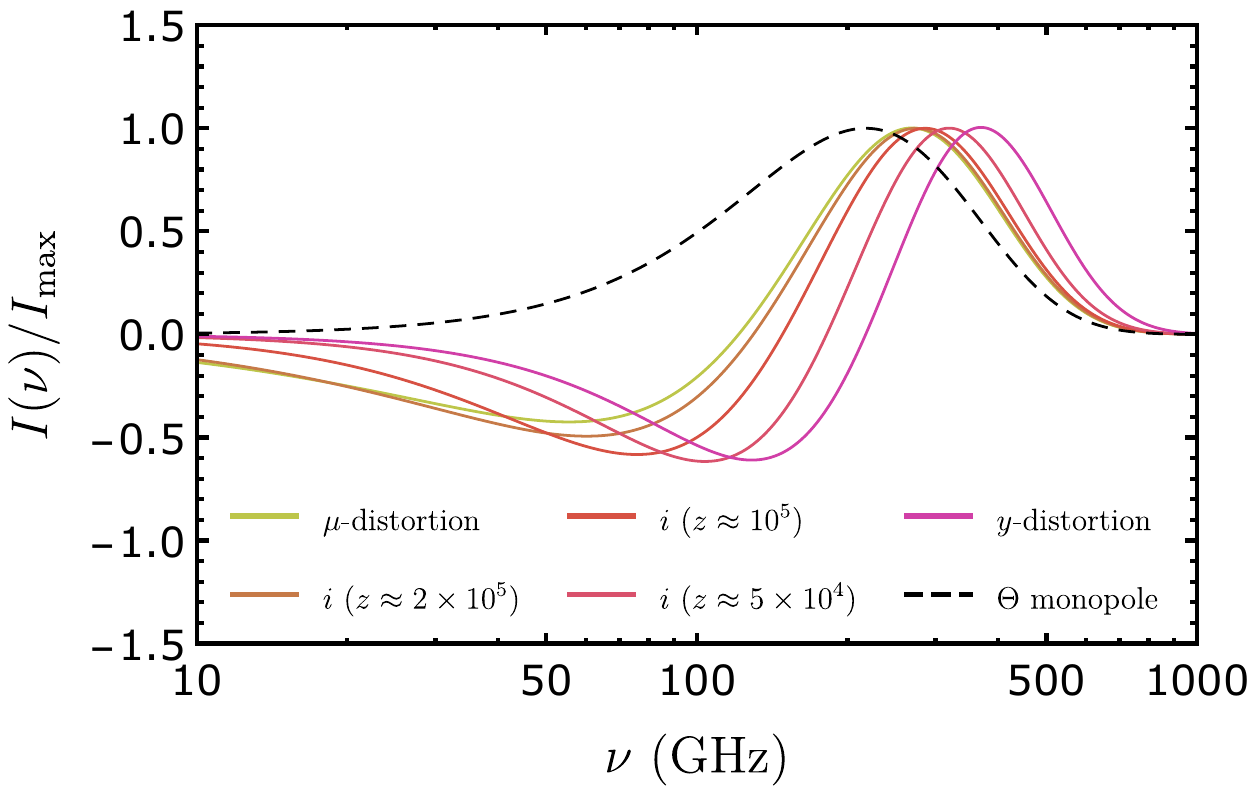}
\caption{\footnotesize{This plot shows the spectral shapes (normalized at the maximum) $I(\nu)$ for $\mu$- and $y$-distortions, together with the spectra for $i$-type distortions at redshifts $z = \mathcal{O}(\num{2d5})$, $z = \mathcal{O}(\num{1d5})$ and $z = \mathcal{O}(\num{5d4})$ \tblack{and the spectral shape of the monopole of temperature anisotropies $\Theta$}. We see that for increasing redshift, the maximum, minimum and zero of the occupation numbers are moved towards lower frequencies.}}
\label{fig:shapes}
\end{figure}

\noindent If we consider \eq{i-general}, we see that we can write down the observed photon spectrum in terms of shapes $\mathcal{I}_a$ and corresponding amplitudes $\mu_a$ where, for example \cite{Zeldovich:1969ff, Sunyaev:1970er}:
\begin{itemize}[leftmargin=*]
\item $a = 1$ corresponds to $\mu$-type occupation number, \ie (recalling that $x\equiv h\nu/k_\tu{B}T$)
\begin{equation}
\label{eq:mu-occupation}
\begin{split}
\mathcal{I}_1 &= \frac{2h\nu^3}{c^2}\frac{e^{x}}{(e^{x} - 1)^2}\bigg(\frac{x}{2.19} - 1\bigg) \\
&\equiv\frac{2h\nu^3}{c^2}\times n^{(\mu)}(\nu)\,\,;
\end{split}
\end{equation}
\item $a = 2$ corresponds to $y$-type occupation number, \ie
\begin{equation}
\label{eq:y-occupation}
\begin{split}
\mathcal{I}_2 &= \frac{2h\nu^3}{c^2}\frac{xe^{x}}{(e^{x} - 1)^2}\bigg[x\bigg(\frac{e^x + 1}{e^x - 1}\bigg) - 4\bigg] \\
&\equiv\frac{2h\nu^3}{c^2}\times n^{(y)}(\nu)\,\,;
\end{split}
\end{equation}
\end{itemize}
and so on. Besides $\mu$-, $i$- and $y$-type distortions, that we have discussed in \sect{distortions-review}, one must also consider the fact that the uniform part of temperature perturbations $\Theta$ is not known a priori and must be fit simultaneously with the spectral distortions: for this reason we also consider the $t$-type occupation number, \ie \cite{Khatri:2013dha}
\begin{equation}
\label{eq:t-occupation}
\begin{split}
\mathcal{I}_t &= \frac{2h\nu^3}{c^2}\frac{xe^{x}}{(e^{x} - 1)^2} \\
&\equiv\frac{2h\nu^3}{c^2}\times n^{(t)}(\nu)\,\,.
\end{split}
\end{equation}
We do not include foregrounds in our analysis since, for PIXIE, the noise penalty for rejecting foregrounds 
is only 2\%, and this noise penalty has been included in all the estimates of CMB sensitivity by the PIXIE collaboration \cite{Kogut:2011xw}. 

We can then write down the signal-to-noise, in terms of amplitudes $\mu_a$ and spectra $\mathcal{I}_a$ as (dropping factors of $2$ for simplicity)
\begin{equation}
\label{eq:StoN}
\bigg(\frac{S}{N}\bigg)^2 = \sum
_c\frac{\Big[\sum_a \mathcal{I}_a(\nu_c)\times(\mu_a - \bar{\mu}_a)\Big]^2}{(\delta I(\nu_c))^2}\,\,,
\end{equation}
where $\bar{\mu}_a$ are the fiducial values of the amplitudes, and $\delta I(\nu_c)$ is the noise at each frequency channel $c$:
\begin{itemize}[leftmargin=*]
\item PIXIE will have $400$ channels ($15\,\mathrm{GHz}$-wide) from $30\,\mathrm{GHz}$ to $6\,\mathrm{THz}$: however, we see from \fig{shapes} that the signals that we consider go quickly to zero beyond $\nu\approx 1000$, so the sum over channels in \eq{StoN} will stop there;
\item $\delta I$ for PIXIE, as from Fig.~12 of \cite{Kogut:2011xw}, is expected to be $\num{5d-26}\,\mathrm{W}\,\mathrm{m}^{-2}\,\mathrm{Hz}^{-1}\,\mathrm{sr}^{-1}$. 
\end{itemize}

If we want to marginalize over some of the amplitudes $\mu_a$ (see \cite{Albrecht:2009ct}, for example), we can use the fact that for a Gaussian with inverse covariance matrix (Fisher matrix) $F$ given by
\begin{equation}
\label{eq:StoN-marginalize-1}
F = 
\begin{pmatrix}
\tilde{F} & S \\
S^T & M
\end{pmatrix}
\,\,,
\end{equation}
where $\tilde{F}$ is the sub-matrix that spans the parameters that we are interested in, the marginalized Fisher matrix will be equal to
\begin{equation}
\label{eq:StoN-marginalize-2}
F_\tu{marg} = \tilde{F} - SM^{-1}S^T\,\,.
\end{equation}
For \eq{StoN}, we will want to marginalize over $t$ and $y$, so $M$ will be the $2\times 2$ matrix
\begin{equation}
\label{eq:StoN-marginalize-3}
M_{ab} = \sum_{c}\frac{\mathcal{I}_a(\nu_c)}{\delta I(\nu_c)}\frac{\mathcal{I}_b(\nu_c)}{\delta I(\nu_c)}\,\,,
\end{equation}
with $a,b = y,t$. Similar expressions can be derived for $S$ and its transpose, while $\tilde{F}$ is simply given by \eq{StoN} with $a$ running on all components except 
$y$ and $t$. 
If we had instead supposed that 
the two 
$y$ and $t$ amplitudes were known, we could just have taken $\tilde{F}$ as Fisher matrix for \eq{StoN}.

In this work we have not considered $i$-distortions, so $F$ will be a $3\times 3$ matrix with $a,b,c = \mu,y,t$: marginalizing over $y$ and $t$ amplitudes, as described in Eqs.~\eqref{eq:StoN-marginalize-2} and \eqref{eq:StoN-marginalize-3}, we obtain $\sigma_{\mu_8} = 1$ for the standard PIXIE configuration. The increments in PIXIE sensitivity that we considered in the text, then, can be 
interpreted 
as 
either an increase in the number $\mathrm{N}$ of frequency channels 
(that would decrease $\sigma_{\mu_8}$ by a factor $\sqrt{\mathrm{N}^\mathrm{PIXIE}/\mathrm{N}^\mathrm{new}}$ ), or a decrease in the instrumental noise $\delta I$ (which instead gives a linear improvement $\delta I^\mathrm{new}/\delta I^\mathrm{PIXIE}$).





\end{document}